\documentclass{aa}
\usepackage{graphicx}
\usepackage{placeins}
\usepackage[varg]{txfonts}
\usepackage{subcaption}
\usepackage{natbib}
\usepackage{hyperref}
\hypersetup{
    colorlinks=true,
    linkcolor=blue,
    citecolor=blue,
    filecolor=magenta,      
    urlcolor=blue,
    bookmarks=true}

\begin{document} 

        \title{Extended far-UV emission surrounding asymptotic giant branch stars as seen by GALEX}
   
        \author{V.~Răstău\inst{\ref{inst1}} \and M.~Me\v{c}ina\inst{\ref{inst1}} \and F.~Kerschbaum\inst{\ref{inst1}} \and 
                                H.~Olofsson\inst{\ref{inst2}} \and M.~Maercker\inst{\ref{inst2}} \and 
                                M.~Drechsler\inst{\ref{inst3}} \and
                                X.~Strottner\inst{\ref{inst4}} \and
                                L.~Mulato\inst{\ref{inst5}}}
                                
        \institute{Department of Astrophysics, University of Vienna, Türkenschanzstrasse 17, 1180 Vienna, Austria \label{inst1}
                                        \and
                                        Department of Space, Earth and Environment, Chalmers University of Technology, Onsala Space Observatory, 439 92 Onsala, Sweden \label{inst2}
                                        \and
                                        Sternwarte Bärenstein, Feldstraße 17, 09471 Bärenstein, Germany \label{inst3}
                                        \and
                                        Montfraze, 01370 Saint Etienne Du Bois, France \label{inst4}
                                        \and
                                        2SPOT (Southern Spectroscopic Observatory Team), 45 Chemin du Lac, 38690 Châbons, France \label{inst5}}
 
        \abstract
        {}
        {Our goal is to study the long-term mass-loss rate characteristics of asymptotic giant branch (AGB) stars through wind--wind and wind--interstellar medium interaction.}
        {Far-ultraviolet (FUV) images from the \textit{Galex} survey are used to investigate extended UV emission associated with AGB stars.}
        {FUV emission was found towards eight objects. The emission displays different shapes and sizes; interaction regions were identified, often with infrared counterparts, but no equivalent near-ultraviolet (NUV) emission was found in most cases.}
        {The FUV emission is likely attributed to shock-excited molecular hydrogen, considering the lack of NUV emission and the large space velocities of the objects, and makes it possible to trace old structures that are too faint to be observed, for instance, in the infrared.}


        \maketitle

\section{Introduction}

\indent\indent The asymptotic giant branch (AGB) phase is the last evolutionary stage of low- to intermediate-mass stars (0.8 < M < 8 M$_{\sun}$) before they become white dwarfs \citep{herwig2005}. Subject to strong convective dredge-up and slow but pronounced dusty winds, AGB stars are major contributors to the enrichment of the interstellar medium (ISM) \citep{schneider2014}. Understanding how the above combination impacts their lifetimes is therefore crucial to our understanding of both stellar evolution and galactic-scale changes.

\par The mass-loss (ML) process of AGB stars is normally viewed as a slow and continuous wind \citep{holof2018} that is responsible for the formation of extended dust and gas structures around such circumstellar envelopes (CSEs). However, more   evidence has emerged in support of more complex scenarios regarding the overall ML process, and that of the formation of extended features \citep{maercker2012, cox2012, decin2020}. For example, the strong thermal pulses that occur in this evolutionary phase produce fast and high ML winds, which then collide with the slower interpulse winds. A shock develops, following this wind--wind interaction, which can also create large structures around AGB stars (i.e. geometrically thin shells, filaments, and clumps). Analysing the morphology of extended features can reveal information about the ML process and the central object, as it did for example in the case of R Sculptoris \citep{maercker2012}. \citet{cox2012} performed an infrared morphological analysis of a large sample of AGB stars showing extended emission, and divided the extended structures into four main categories based on their shape: fermata (large arcs), eyes (concentric arcs), rings (circular structures), and irregular (for diffuse structures or any structures that do not fall into the previous categories). Our discussion here follows this classification.

\par To date only a handful of AGB stars showing extended emission in the ultraviolet (UV) have been studied, the most famous one being \object{Mira} \citep[see][]{martin2007}. Along with the features found around \object{CW Leo} and \object{RW LMi} \citep[see][]{sahai2010, sahai2014} these are fine examples of wind--ISM interaction. Recently, \citet{sanchez2015} detected, for the first time, UV emission cospatial with a detached dust shell previously found in the infrared (IR), which was attributed to outflow and/or ISM shocks. All in all, not much is known about AGB winds in this wavelength regime. Therefore, in the scope of this project our aim is to expand the sample of AGB stars that display extended UV emission, more specifically far-UV (FUV) emission (0.13 – 0.18 $\mu$m) taken from the GALEX catalogue. In section \ref{s2} we briefly describe the sample and the various methods used. A more detailed analysis is provided for each object in Sect.~\ref{s3}, while Sect.~\ref{s4} is reserved for the discussion of our results and the conclusions we draw from them.

\section{Sample and method} \label{s2}

\begin{figure*}[!]
\centering
        \begin{subfigure}{.24\linewidth}
                \includegraphics[width = \linewidth]{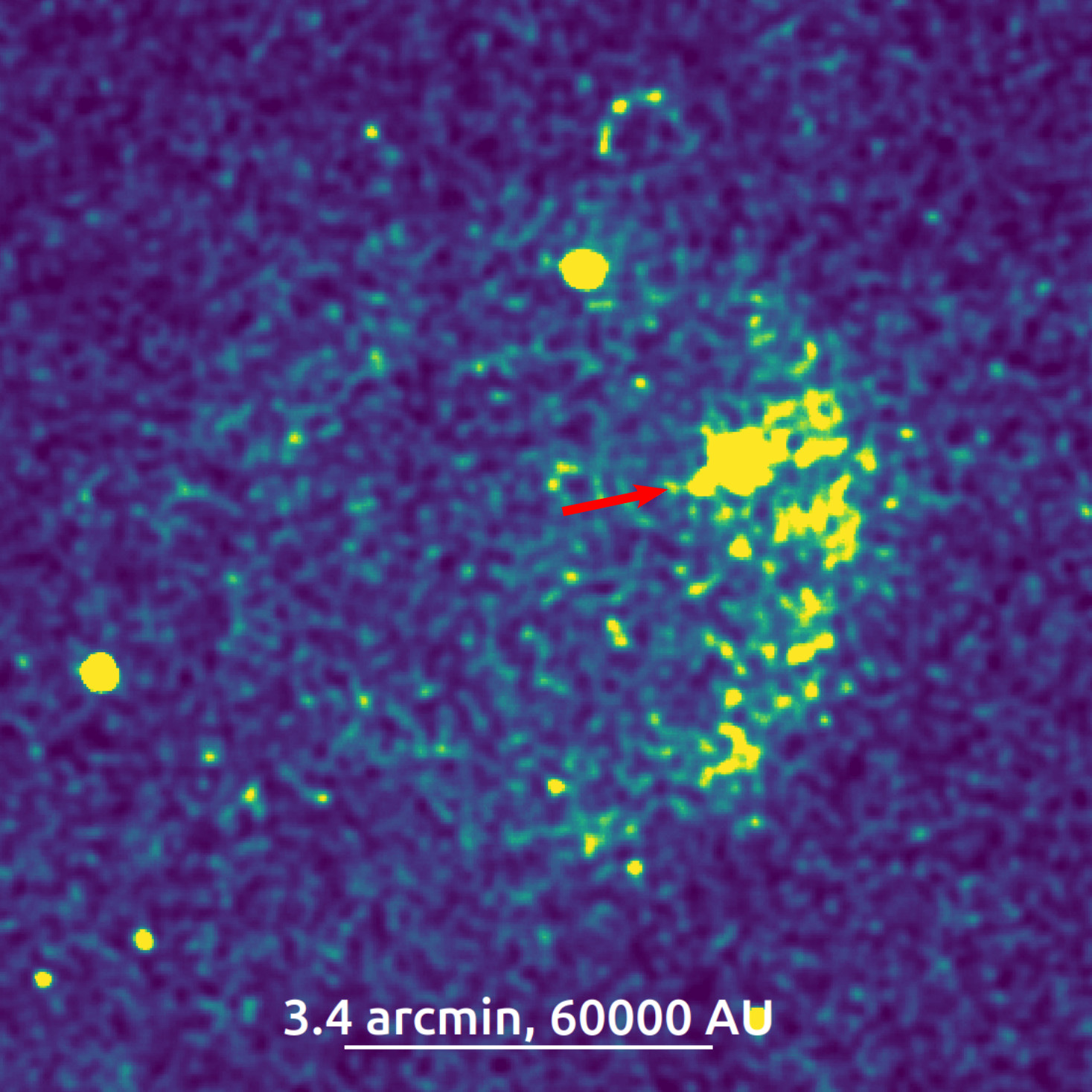}
                \caption{\object{U Ant}} \label{fuvalla}
        \end{subfigure} 
        \begin{subfigure}{.24\linewidth}
                \includegraphics[width = \linewidth]{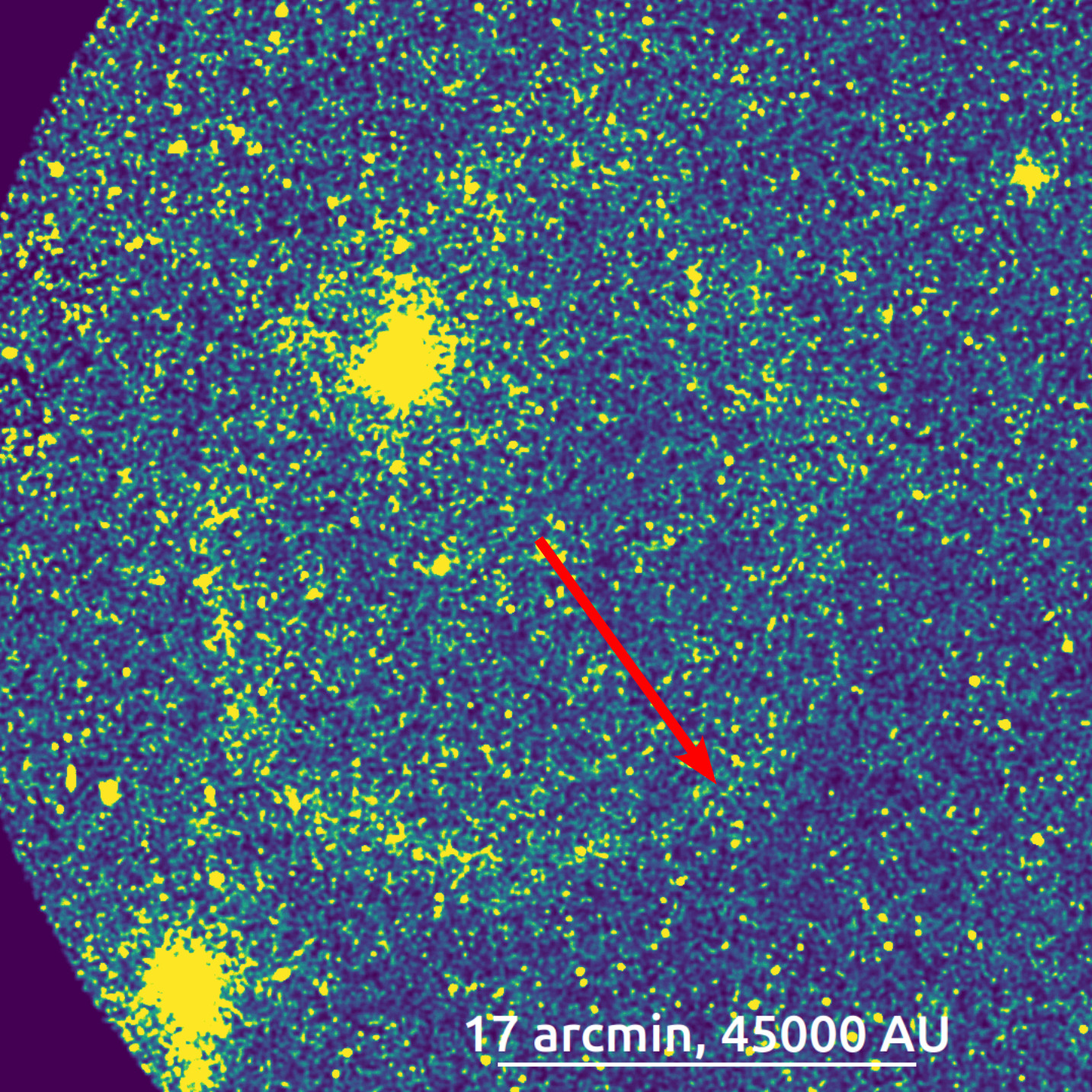}
                \caption{\object{R Dor}} \label{fuvallb}
        \end{subfigure}
        \begin{subfigure}{.24\linewidth}
                \includegraphics[width = \linewidth]{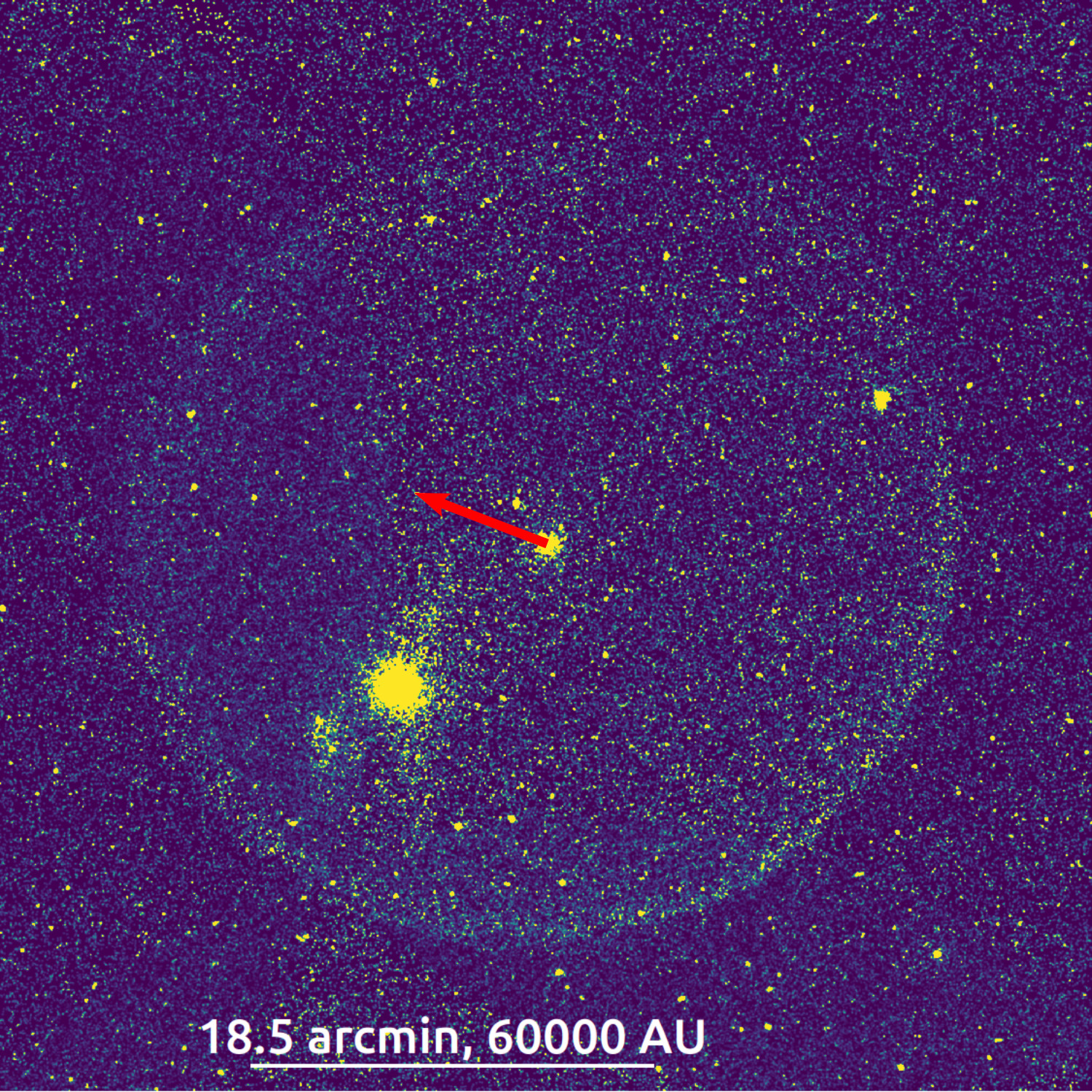}
                \caption{\object{$\beta$ Gru}} \label{fuvallc}
        \end{subfigure}
        \begin{subfigure}{.24\linewidth}
                \includegraphics[width = \linewidth]{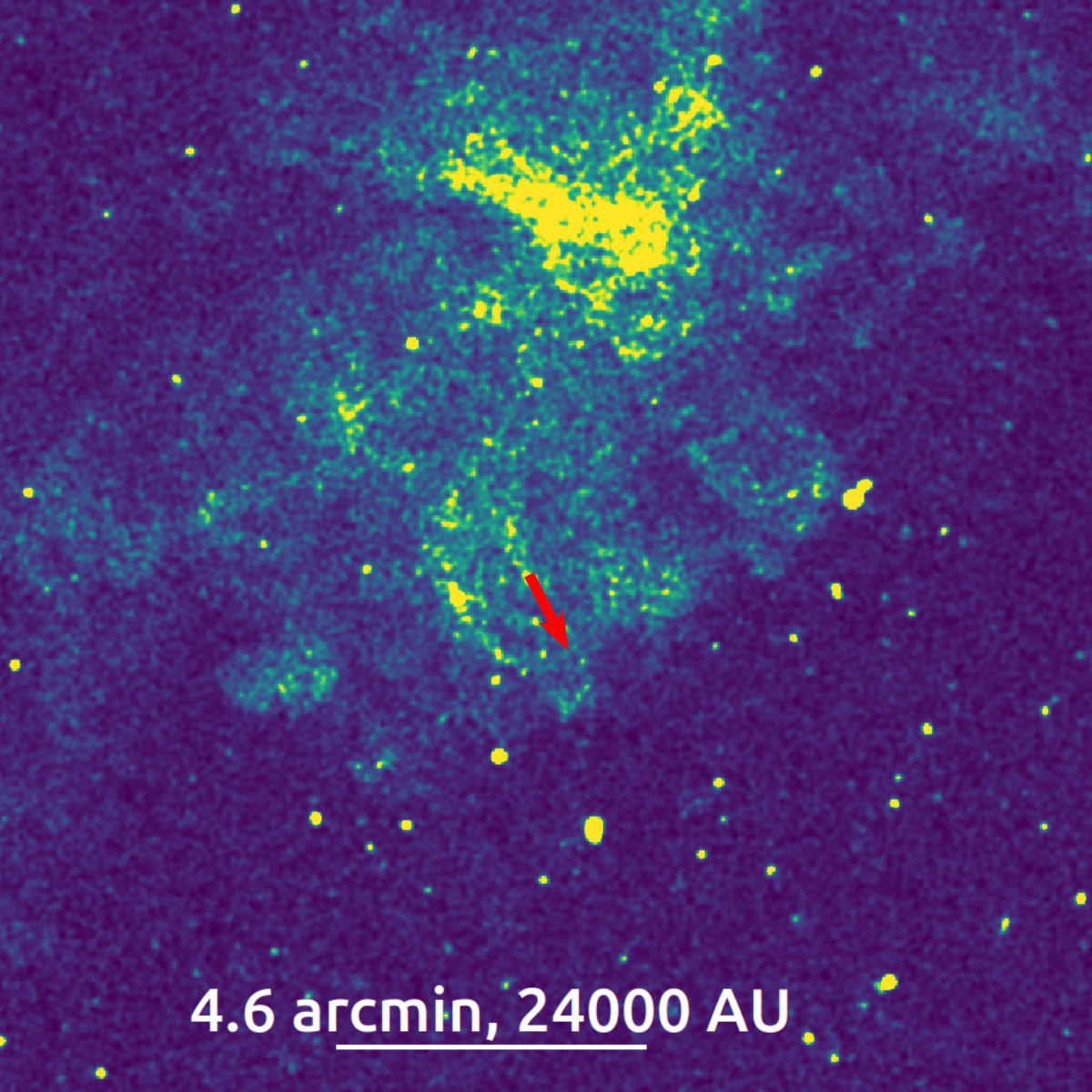}
                \caption{\object{W Hya}} \label{fuvalld}
        \end{subfigure}
\end{figure*}
\begin{figure*}
\centering
\ContinuedFloat
        \begin{subfigure}{.24\linewidth}
                \includegraphics[width = \linewidth]{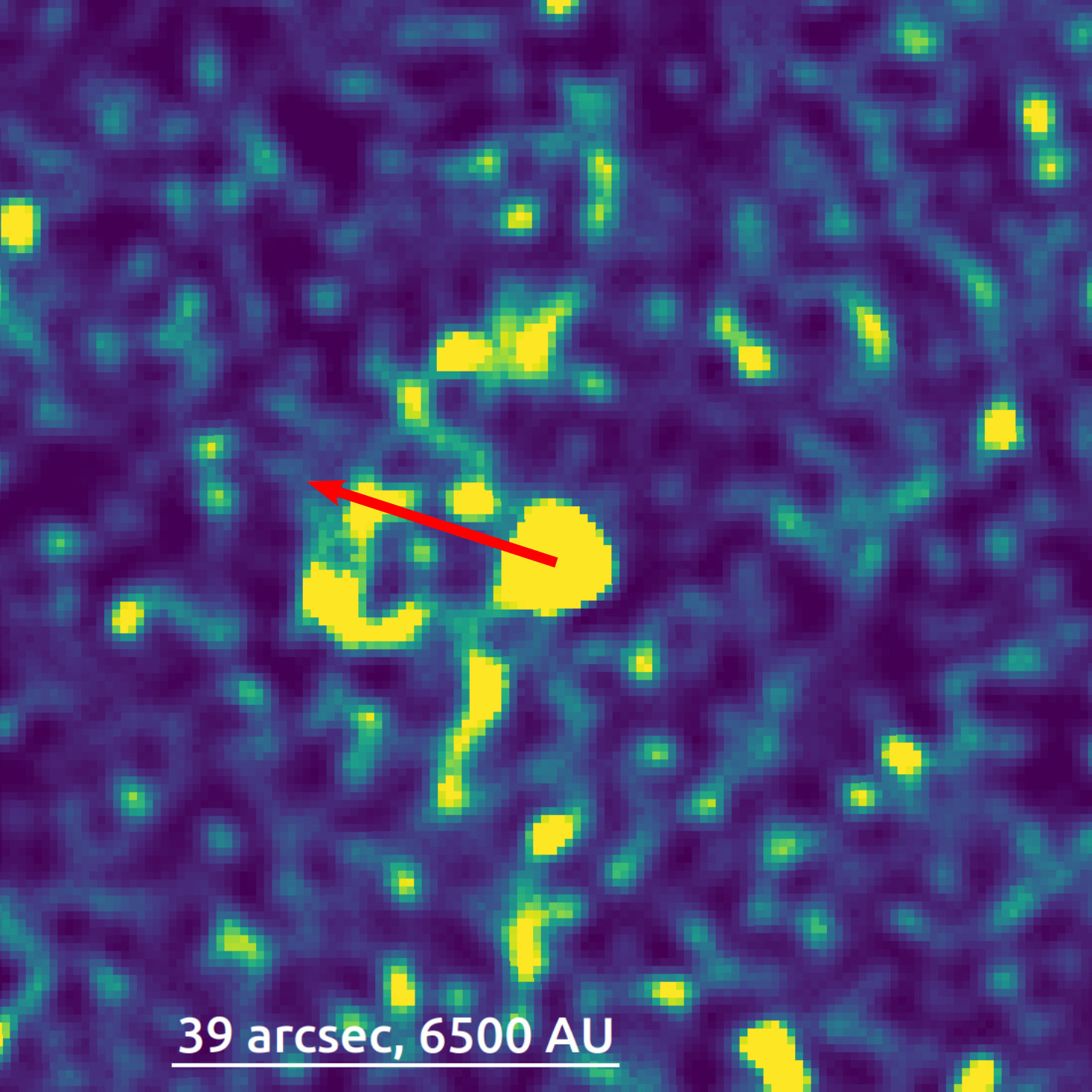}
                \caption{\object{TV Psc}} \label{fuvalle}
        \end{subfigure}
        \begin{subfigure}{.24\linewidth}
                \includegraphics[width = \linewidth]{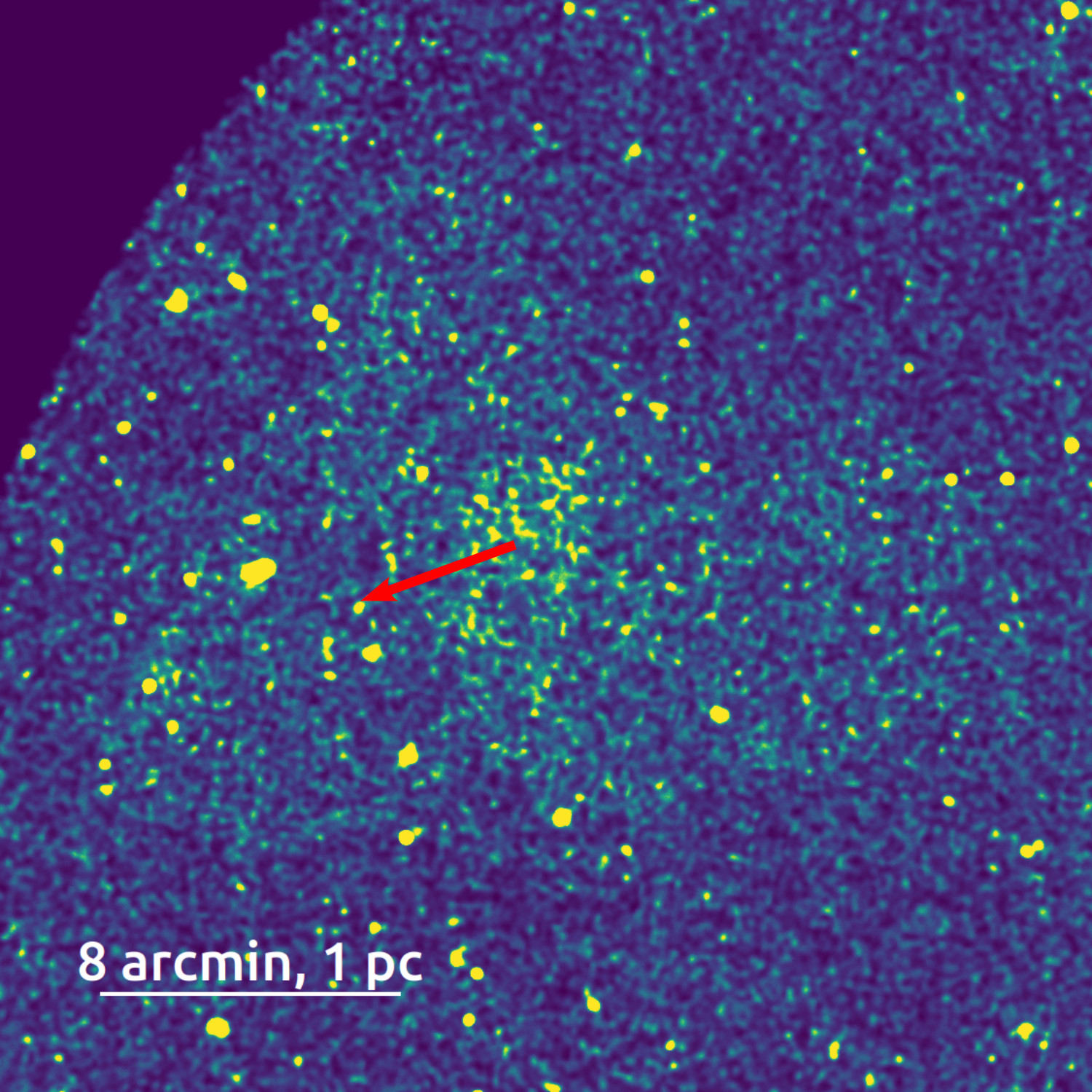}
                \caption{\object{RZ Sgr}} \label{fuvallf}
        \end{subfigure}
        \begin{subfigure}{.24\linewidth}
                \includegraphics[width = \linewidth]{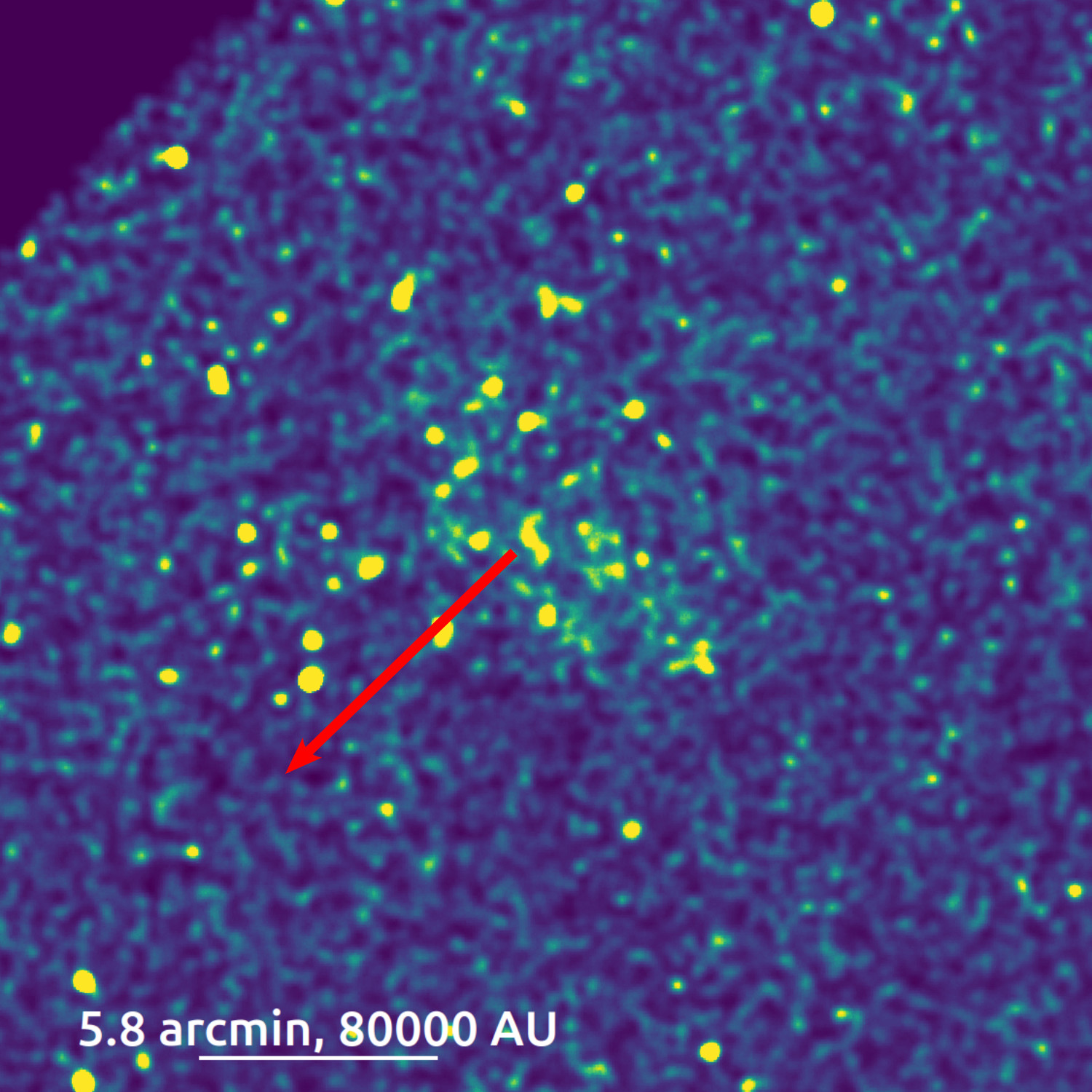}
                \caption{\object{DM Tuc}} \label{fuvallg}
        \end{subfigure}
        \begin{subfigure}{.24\linewidth}
                \includegraphics[width = \linewidth]{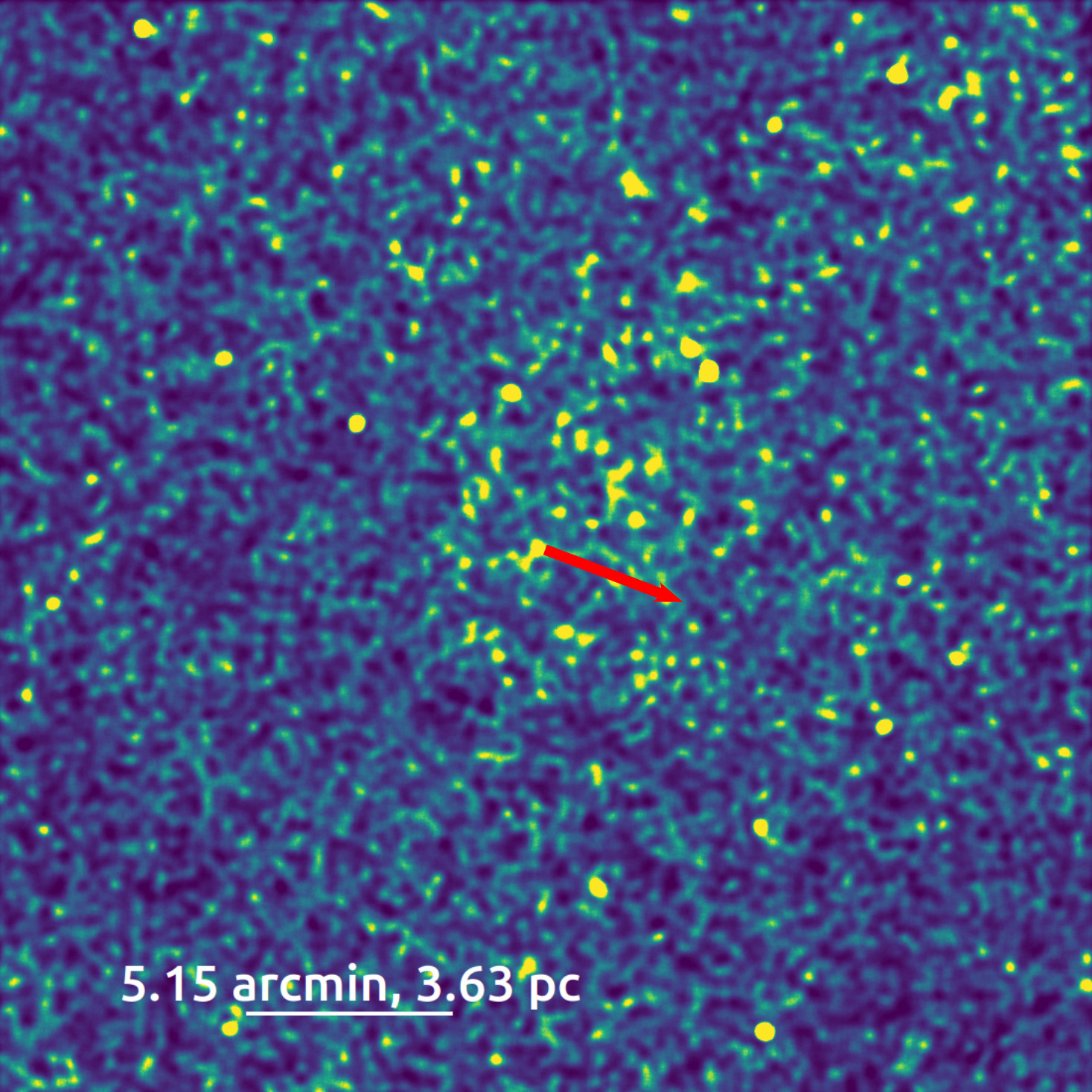}
                \caption{\object{V420 Vul}} \label{fuvallh}
        \end{subfigure}
\caption{GALEX catalogue FUV images of the stars and their surroundings. North is up and east to the left; the same orientation is used for all the images. The arrows (originating at the location of each star) represent the projected distance the stars would travel over a timescale of 5000 years, with the exception of \object{TV Psc} (where the distance corresponds to 500 years), \object{RZ Sgr} (40\,000 years), and \object{V420 Vul} (20\,000 years). The space motion vector indicates the velocity relative to the local standard of rest (LSR) \citep{schonrich2010} and is derived from proper motion and radial velocity data using the Astropy coordinates package. The large bright spots seen in the images for \object{$\beta$ Gru}, \object{R Dor,} and \object{U Ant} come from objects along the line of sight, either background stars or galaxies.} \label{fuvall} 
\end{figure*}

\begin{figure}[ht]
\centering
        \includegraphics[width=\linewidth]{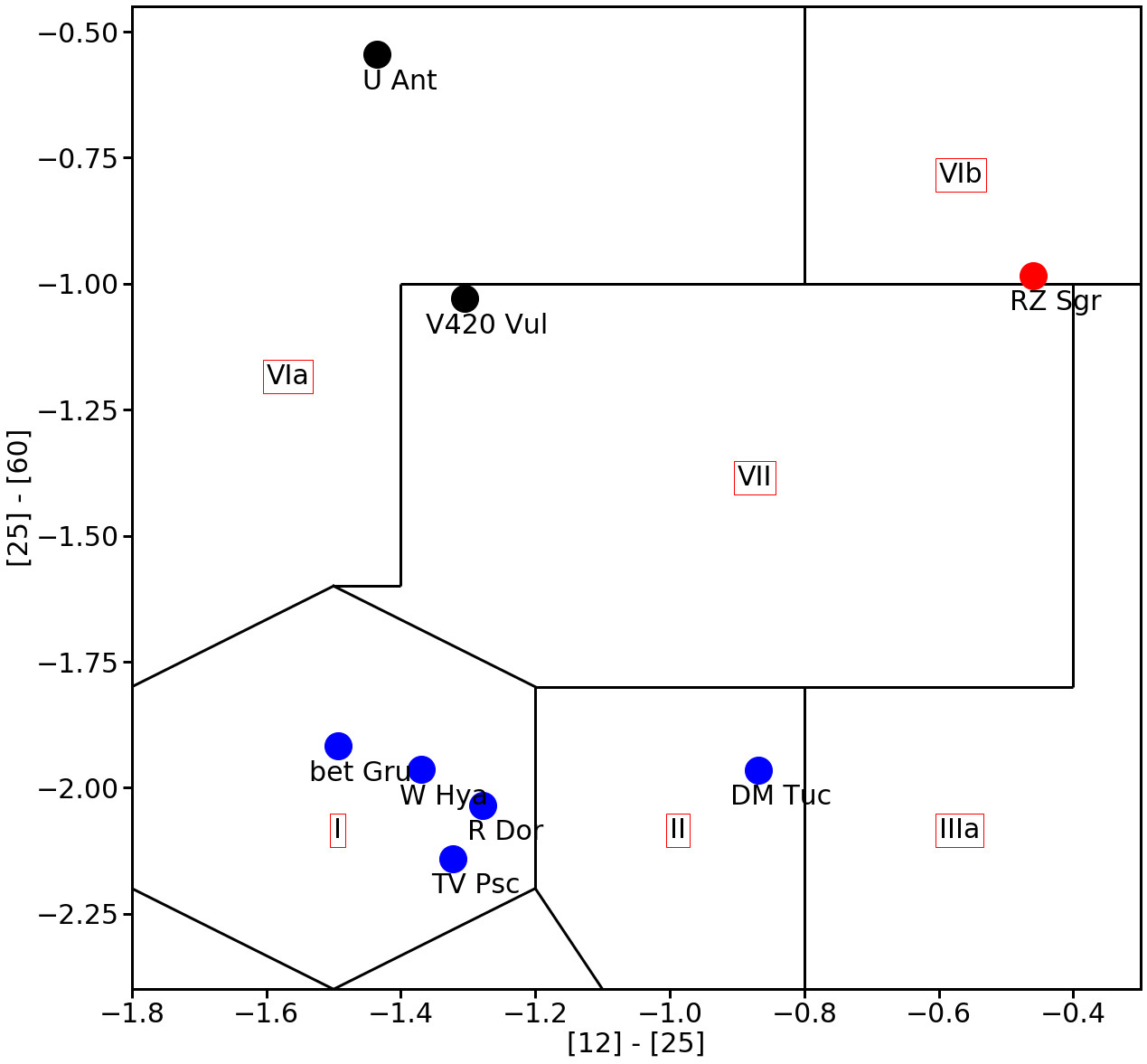}
\caption{IRAS colour--colour diagram \citep{vdv1988}. The objects are colour-coded as follows: blue dots for M-type stars, red dots for S-type stars, and black dots for C-type stars.} \label{vdvdiag} 
\end{figure}

\begin{table*}
\caption{Stellar parameters} \label{params}
\centering
\resizebox{\linewidth}{!}{
\begin{tabular}{l*{13}l}
                                                                        \hline\hline
                            & RA 
                & Dec 
                & $\pi$ 
                & Distance
                & $T_{\mathrm{eff}}$
                & $L$ 
                & $P$
                & $\dot{M}$ 
                & Var type 
                & Spectral type \\ 
                &
                &
                & [mas]
                & [pc]
                & [K]
                & [$\mathrm{L_{\sun}}$]
                & [days] 
                & [$\mathrm{M_\odot}$\,yr$^{-1}$]
                &
                & \\ \hline

\object{U Ant}          & 10$^{\rm h}$35$^{\rm m}$12.85$^{\rm s}$ 
                        & -39${\degr}$33${\arcmin}$45.32${\arcsec}$
                                        & 3.4\tablefootmark{\,a}  
                        & 294\tablefootmark{\,a} 
                        & 2300\tablefootmark{\,d} 
                        & 4500\tablefootmark{\,d} 
                        & -- 
                        & $4.0\,\times\,10^{-8}$\tablefootmark{\,m}
                        & Lb 
                        & C-N3 \\

\object{R Dor}                  & 04$^{\rm h}$36$^{\rm m}$45.59$^{\rm s}$ 
                        & -62${\degr}$04${\arcmin}$37.79${\arcsec}$
                        & 22.7\tablefootmark{\,a} 
                        & 44\tablefootmark{\,a}  
                        & 2710\tablefootmark{\,e} 
                        & 2800\tablefootmark{\,e} 
                        & 223\tablefootmark{\,k} 
                        & $1.6\,\times\,10^{-7}$\tablefootmark{\,n}
                        & SRb 
                        & M8III:e \\
                                                                        
\object{$\beta$ Gru}    & 22$^{\rm h}$42$^{\rm m}$40.05$^{\rm s}$ 
                        & -46${\degr}$53${\arcmin}$04.47${\arcsec}$
                                        & 18.4\tablefootmark{\,b} 
                        & 54 & 3480\tablefootmark{\,f,g} 
                        & 2500\tablefootmark{\,f} 
                        & 37\tablefootmark{\,k}
                        & $1.0\,\times\,10^{-9}$ 
                        & SRb 
                        & M4.5III \\

\object{W Hya}                  & 13$^{\rm h}$49$^{\rm m}$02.00$^{\rm s}$ 
                        & -28${\degr}$22${\arcmin}$03.53${\arcsec}$                                                                                             
                        & 11.0\tablefootmark{\,a}
                        & 87\tablefootmark{\,a} 
                        & 2600\tablefootmark{\,h} 
                        & 4730\tablefootmark{\,h} 
                        & 390\tablefootmark{\,l} 
                        & $1.5\,\times\,10^{-7}$\tablefootmark{\,o}
                        & SRa 
                        & M7.5-9e \\

\object{TV Psc}                 & 00$^{\rm h}$28$^{\rm m}$02.91$^{\rm s}$ 
                        & +17${\degr}$53${\arcmin}$35.25${\arcsec}$
                                        & 6.0\tablefootmark{\,c} 
                        & 167 
                        & 3500* 
                        & 2500* 
                        & 49\tablefootmark{\,k}
                        & $1.0\,\times\,10^{-8}$ 
                        & SR 
                        & M3III \\

\object{RZ Sgr}                 & 20$^{\rm h}$15$^{\rm m}$28.40$^{\rm s}$ 
                        & -44${\degr}$24${\arcmin}$37.47${\arcsec}$
                                        & 2.3\tablefootmark{\,a} 
                        & 432\tablefootmark{\,a} 
                        & 2400\tablefootmark{\,i} 
                        & 5000\tablefootmark{\,j} 
                        & 172\,/\,338\tablefootmark{\,k} 
                        & $3.0\,\times\,10^{-6}$\tablefootmark{\,p}
                        & M 
                        & S4,4ep \\
                                                                        
\object{DM Tuc}                 & 22$^{\rm h}$57$^{\rm m}$05.85$^{\rm s}$ 
                        & -57${\degr}$24${\arcmin}$04.22${\arcsec}$ 
                                        & 4.3\tablefootmark{\,c} 
                        & 230 
                        & 2700* 
                        & 3340\tablefootmark{\,j} 
                        & 75\,/\,145\tablefootmark{\,k} 
                        & $4.1\,\times\,10^{-7}$\tablefootmark{\,j}
                        & SRb 
                        & M8III \\

\object{V420 Vul}               & 20$^{\rm h}$59$^{\rm m}$36.83$^{\rm s}$ 
                        & +26${\degr}$28${\arcmin}$34.42${\arcsec}$
                                        & 0.4\tablefootmark{\,c} 
                        & 2429 
                        & 3000* 
                        & 4000* 
                        & 377\tablefootmark{\,k}
                        & $1.7\,\times\,10^{-7}$ 
                        & M 
                        & N:e \\ \hline
\end{tabular}
}
\tablefoot{Due to a lack of literature data, the luminosity and effective temperature values flagged with asterisks (*) were calculated by us using the radiative transfer code \textit{More of DUSTY} \citep{groenewegendusty}. The luminosities for U Ant, RZ Sgr, R Dor, and W Hya were scaled based on the results of the various works referenced in this table and the distances derived by \citet{miora2022}. \\
\textbf{References.} \tablefoottext{a}{\citet{miora2022}}
\tablefoottext{b}{\citet{vanleeuwen2007}} 
\tablefoottext{c}{\citet{gaiadr3}} 
\tablefoottext{d}{\citet{maercker2018}}
\tablefoottext{e}{\citet{ohnaka2019}}
\tablefoottext{f}{\citet{judge1986}} 
\tablefoottext{g}{\citet{engelke2006}}
\tablefoottext{h}{\citet{ramol2014}}
\tablefoottext{i}{\citet{schoier2013}} 
\tablefoottext{j}{\citet{winters2003}} 
\tablefoottext{k}{\citet{gcvs51}}
\tablefoottext{l}{\citet{vogt2016}}
\tablefoottext{m}{\citet{kerschbaum2017}}
\tablefoottext{n}{\citet{maercker2016}}
\tablefoottext{o}{\citet{khouri2014}} 
\tablefoottext{p}{\citet{ramstedt2009}}}
\end{table*}

\begin{table}[ht] 
\caption{Stellar motion data} \label{motion}
\centering
\resizebox{\linewidth}{!}{
\begin{tabular}{l*{6}l}                 \hline\hline
                                & $pm_{\mathrm{RA}}$
                & $pm_{\mathrm{DEC}}$
                & RUWE
                & $r_v$
                & $V_{\mathrm{s_{lsr}}}$        \\
                & $\mathrm{[mas\,yr^{-1}]}$ 
                & $\mathrm{[mas\,yr^{-1}]}$  
                &
                & $\mathrm{[km\,s^{-1}]}$
                & $\mathrm{[km\,s^{-1}]}$       \\ \hline
\object{U Ant}                                                  & -31.5 \tablefootmark{\,a}
                                        &  2.6 \tablefootmark{\,a}
                                        & --
                                        & 41 \tablefootmark{\,d}     
                                        & 42                          \\
\object{R Dor}                                                  & -69.4 \tablefootmark{\,b} 
                                        & -75.8 \tablefootmark{\,b}
                                        & --     
                                        & 26.1 \tablefootmark{\,d}   
                                        & 40      \\
\object{$\beta$ Gru}                                & 135.2 \tablefootmark{\,b}   
                                        & -4.4 \tablefootmark{\,b}   
                                        & --     
                                        & -0.3 \tablefootmark{\,d}  
                                        & 20    \\
\object{W Hya}                                                  & -51.8 \tablefootmark{\,c} 
                                        & -59.7 \tablefootmark{\,c} 
                                        & --     
                                        & 42.3 \tablefootmark{\,d}  
                                        & 50    \\
\object{TV Psc}                                 & 114.6 \tablefootmark{\,a}
                                        &  20.3 \tablefootmark{\,a}
                                        & 2.4404 \tablefootmark{\,a}
                                        & 5.31 \tablefootmark{\,e}  
                                        & 78      \\
\object{RZ Sgr}                             & 15.5 \tablefootmark{\,a}
                                        & -10.9 \tablefootmark{\,a}
                                        & --
                                        & -38.25 \tablefootmark{\,a}
                                        & 41    \\
\object{DM Tuc}                                                 & 65.8 \tablefootmark{\,a} 
                                        & -55.9 \tablefootmark{\,a}
                                        & 1.1879 \tablefootmark{\,a}
                                        & -35.5 \tablefootmark{\,f} 
                                        & 87    \\
\object{V420 Vul}                                               & -9.4 \tablefootmark{\,a}  
                                        & -4.1 \tablefootmark{\,a}  
                                        & 1.8602 \tablefootmark{\,a} 
                                        & 14.9 \tablefootmark{\,a}   
                                        & 132   \\ \hline               
\end{tabular}
}
\tablefoot{Proper motions ($pm_{\mathrm{RA}}$, $pm_{\mathrm{DEC}}$), Gaia renormalised unit weight error (RUWE), radial velocity ($r_v$) and space velocity with respect to the local standard of rest  ($V_{\mathrm{s_{lsr}}}$, LSR transformation based on \citealt{schonrich2010}). \\
\textbf{References.} \tablefoottext{a}{\citet{gaiadr3}} 
\tablefoottext{b}{\citet{vanleeuwen2007}}
\tablefoottext{c}{\citet{gaiadr2}}
\tablefoottext{d}{\citet{hiprvgontch}} 
\tablefoottext{e}{\citet{hiprvfam}}
\tablefoottext{f}{The value is based on the result of \citet{winters2003} and the LSR transformation of \citet{schonrich2010}}}
\end{table}

\begin{table}[ht] 
\caption{GALEX observations} \label{galobs}
\centering
\resizebox{\linewidth}{!}{
\begin{tabular}{l*{3}l}                   \hline\hline
            & \multicolumn{2}{l}{Exposure time [s]}
            & Observation date          \\ 
            & NUV
            & FUV
            & 
            \\ \hline
U Ant       & 6379          & 1119            & 7,11 Apr 2009             \\
R Dor       & 206           & 206             & 28 Sep 2008               \\
$\beta$ Gru & 199, 215, 244 & 199, 215, 244   & 8 Aug 2007                \\
W Hya       & 4420          & 4420            & 16 May 2009               \\
TV Psc      & 460           & 339             & 28 Oct 2003, 26 Sep 2004  \\
RZ Sgr      & 210           & 210             & 4 Jul 2007                \\
DM Tuc      & 385           & 320             & 9 Aug 2003                \\
V420 Vul    & 398           & 255             & 30 Jun 2004               \\ \hline
\end{tabular}
}
\tablefoot{Exposure times and observation dates of the GALEX data used for this study (GALEX archive -- GR 6/7).} 
\end{table}

\begin{table}[ht] 
\caption{Results on the extended emission} \label{fluxes}
\centering
\resizebox{\linewidth}{!}{
\begin{tabular}{l*{6}l}                 \hline\hline
                & $F_{\mathrm{ex}}$
                                & $F_{\mathrm{fuv}}$            
                & $R_{\mathrm{es}}$     
                & $t_{\mathrm{kin}}$                                                              \\
                & [mJy]
                & [mJy]
                & [AU]
                & [yrs]                                                                           \\ \hline
\object{U Ant}                                                  & 1.3  & --                     & 50\,000  & 23\,700      \\
\object{R Dor}                                                  & 12.7 & --                     & 37\,000  & 17\,500      \\
\object{$\beta$ Gru}                                & 9.7  & 3.83                   & 58\,000  & 27\,500      \\
\object{W Hya}                                                  & 0.6  & --                     & 6\,250   & 3\,000       \\
\object{TV Psc}                                 & 0.07 & 0.174                  & 6\,600   & 3\,100       \\
\object{RZ Sgr}                             & 1.6* & --                     & 180\,000 & 85\,000      \\
\object{DM Tuc}                                                 & 1.1  & $3.53\,\times 10^{-3}$ & 10\,000  & --            \\
\object{V420 Vul}                                               & 3.7  & --                     & --       & --            \\ \hline
\end{tabular}
}
\tablefoot{ FUV flux densities ($F_{\mathrm{ex}}$), radius ($R_{\mathrm{es}}$), and kinematic age ($t_{\mathrm{kin}}$) of the extended structures presented in Sect.~\ref{s3}. $\mathrm{F_{fuv}}$ is the GALEX FUV flux of the stars \citep{galex}. We assumed an expansion velocity of the outflow of $\mathrm{10\,km\,s^{-1}}$ for all objects. * The FUV flux density of \object{RZ Sgr} used here corresponds only to the inner, more pronounced, emission.}
\end{table}

\subsection{Sample}
\indent\indent This study focuses on a small sample of AGB stars that show extended FUV emission (0.13 – 0.18 $\mu$m) in the GALEX catalogue \citep{galex}. The sample was obtained following a systematic inspection by eye of GALEX FUV images, carried out via Aladin on the full-sky survey (GALEX GR6/7 FUV / GR6 AIS FUV), with the scope of finding previously unreported sources that display faint extended emission in this wavelength regime, regardless of source type. Any detections were correlated with several catalogues, and those that were associated with known objects were flagged. Here we report on newly found emission associated with variable stars of Mira-type, Semiregular or Irregular (M, SRa, SRb, SR, or Lb), that are listed in the General Catalogue of Variable Stars \citep[GCVS,][]{gcvs51}. Other results, for example on supernova remnants, can be found in \citet{fresen2021} or \citet{kimeswenger2021}. Using the GCVS in the identification process introduces a bias towards AGB stars which are more optically bright and are located towards the low end of the ML spectrum. Our object list consists of five M-type stars (oxygen-rich, C/O < 1), two C-type stars (carbon-rich, C/O > 1), and one S-type star (C/O $\approx$ 1). The stars along with their basic parameters (gathered from the literature where available) are described in Table~\ref{params}. The distances used here are from \citet{miora2022}, derived from the Gaia DR3 parallaxes \citep{gaiadr3}, or derived from the Hipparcos parallaxes \citep{vanleeuwen2007} for the cases where Gaia measurements were not available. Only three objects have a renormalised unit weight error (RUWE) for their Gaia measurements (see Table~\ref{motion}). One of them (TV Psc) has a value that would indicate somewhat poor quality, but since the measurements were not that far off from previous estimates we decided to keep the values from Gaia DR3. We adopted CO mass-loss rates (MLRs) from the literature where available and for the remaining cases we estimated the present-day MLR via spectral energy distribution (SED) fitting done with the \textit{More of DUSTY} code \citep[with an assumed gas-to-dust mass ratio of 200;][]{groenewegendusty}. The fitting was done using model spectra \citep{comarcs2016, comarcs2019} and the gathered stellar parameters (Table~\ref{params}).

\subsection{UV data}
\indent\indent The FUV (bandpass 1344-1786 Å) and near-ultraviolet (NUV, bandpass 1771-2831 Å) data used in this study come from pipeline-calibrated images (in units of counts per second per pixel) retrieved from the GALEX archive, GR6/7\footnote{The data are from \href{https://galex.stsci.edu/GR6/}{https://galex.stsci.edu/GR6/}. U Ant data were calibrated using the GR7 pipeline (\href{http://www.galex.caltech.edu/wiki/Public:Documentation/Chapter_8}{http://www.galex.caltech.edu/wiki/Public:Documentation/Chapter\_8}), while the image data for the rest of the objects were calibrated with the GR6 pipeline (\href{https://galex.stsci.edu/Doc/GI_Doc_Ops7.pdf}{https://galex.stsci.edu/Doc/GI\_Doc\_Ops7.pdf}).}, with a 1.2 degree field of view, a pixel scale of 1.5\arcsec, and a FWHM angular resolution of 4\arcsec\ (FUV) and 5.6\arcsec\ (NUV). Table~\ref{galobs} lists the exposure times of the retrieved images. The emission detected around each star is displayed in Fig.~\ref{fuvall} and throughout Sect.~\ref{s3}. We  also performed some basic FUV flux density estimations for the extended structures (with flux densities obtained via a conversion of the photon counts; see Appendix~\ref{appA}), the values being listed in Table~\ref{fluxes} along with their estimated sizes. The size measurements are based on the absolute distances and the apparent angular sizes of the objects, while the age determination of the extended structures also includes the assumption of a constant expansion velocity of $\mathrm{10\,km\,s^{-1}}$ or other values in those cases where more accurate velocity measurements are available.

\par As can be clearly seen, the FUV emission produces various types of structures, from well-defined rings (\object{$\beta$ Gru,} \object{R Dor}) to faint arcs (eye, fermata: \object{TV Psc}, \object{W Hya}, \object{U Ant}), and to fuzzy faint blobs (irregular: \object{DM Tuc}, \object{RZ Sgr,} \object{V420 Vul}). One important aspect is that no near-UV (NUV, 0.17 -- 0.30 $\mu$m) counterpart was detected for most structures. The exceptions are \object{TV Psc}, which displays some stronger NUV emission around the star, and \object{RZ Sgr}. As discussed in previous studies \citep{martin2007, sahai2010, sahai2014} the presence of FUV emission combined with the lack of NUV emission is indicative of the origin of the FUV flux. One cause could be wind--ISM interaction, where the emission is attributed to shocked molecular hydrogen, which does not emit in the NUV. \citet{sanchez2015} argue that if enough $\mathrm{H_2}$ molecules survive dissociation by the interstellar radiation field, this could show up as clumps of emission in the FUV regime. The FUV data were also compared to far-infrared (FIR) pipeline-calibrated images from the Herschel or Akari surveys. The FUV emission was found to be cospatial with IR emission, with possible implications being discussed in the following sections.

\par Using infrared photometric data, we plotted the objects in an IRAS colour--colour diagram (see Fig.~\ref{vdvdiag}), which was introduced by \citet{vdv1988} to illustrate the evolution of O-rich AGB stars with circumstellar envelopes (CSEs). The different regions denoted by roman numerals lay out the evolutionary sequence as follows: it begins with mostly non-variable early-AGB stars (M-type) that show no evidence of having a CSE (region I), continues through regions II--IV as the stars increase in variability and ML and start developing thicker O-rich CSEs, and ends in regions IV and V where AGB stars that are close to the end of the ML process and those that have reached the planetary nebula stage are found. Regions VI and VII in the diagram contain C- and S-type AGB stars, objects that have seen a shift towards a more C-rich chemistry via dredge-up events caused by thermal pulses, with similar variability and CSE characteristics seen in regions II and III. Finally, group VIII represents outliers and extreme cases of objects seen towards the end of AGB evolution (i.e. objects from regions IV and V). The objects in our sample seem to end up in the expected categories, with the M-type stars in groups specific for stars with no or tenuous CSEs. The C-type stars in the sample ended up in categories containing objects with C-rich CSEs. The S-type star is located where objects that might experience episodic ML are expected, having a colder dust component that could be located further away from the object.

\subsection{CO line observations for $\beta$ Gruis}
\indent\indent The star $\beta$ Gru is the only object that clearly shows an essentially circular FUV shell centred on the star. This warranted an attempt to study the mass-loss history of the object using CO rotational line emission. Observations of the CO($J$\,=\,2–1) line were obtained using the APEX telescope \citep{apex} in April 2022.  No CO($J$\,=\,2–1) emission was detected, having achieved an upper limit of $\mathrm{1\,mK}$ for the rms noise level (in main beam brightness scale; beam width = 27\arcsec ) at a velocity resolution of 2 $\mathrm{km\,s^{-1}}$.

\section{Object analysis} \label{s3}

\subsection{U Antilae}
\indent\indent \object{U Ant} is a C-type irregular variable located 294 pc away. Based on its effective temperature and luminosity, $T_{\mathrm{eff}} \mathrm{= 2300\,K}$ and $L \mathrm{\,= 4500\, L_{\sun}}$ \citep{maercker2018}, it is a relatively evolved AGB star. A low present-day MLR, $\mathrm{4\,\times\,10^{-8}\,M_{\sun}\,yr^{-1}}$, has been estimated using CO line data \citep{kerschbaum2017}; the rough estimate we computed is of the same order of magnitude.

\par
\begin{figure}[htb]
\centering
        \begin{subfigure}{.49\linewidth}
                \includegraphics[width = \linewidth]{uant_fuvang.jpeg}
                \caption{Galex FUV image} \label{uauv}
        \end{subfigure}
        \begin{subfigure}{.49\linewidth}
                \includegraphics[width = \linewidth]{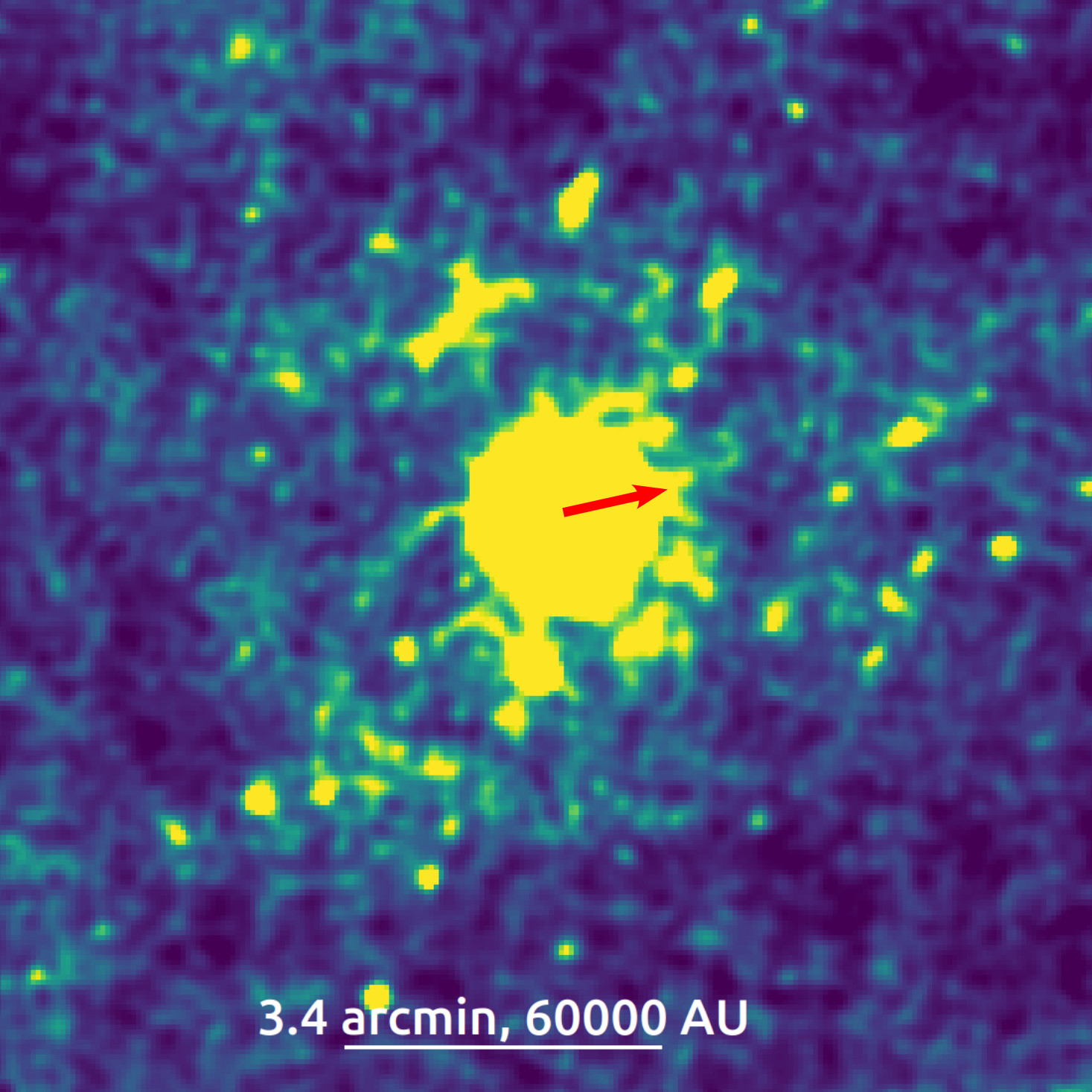}
                \caption{Herschel PACS160 image} \label{uair}
        \end{subfigure}
        \begin{subfigure}{.49\linewidth}
                \includegraphics[width = \linewidth]{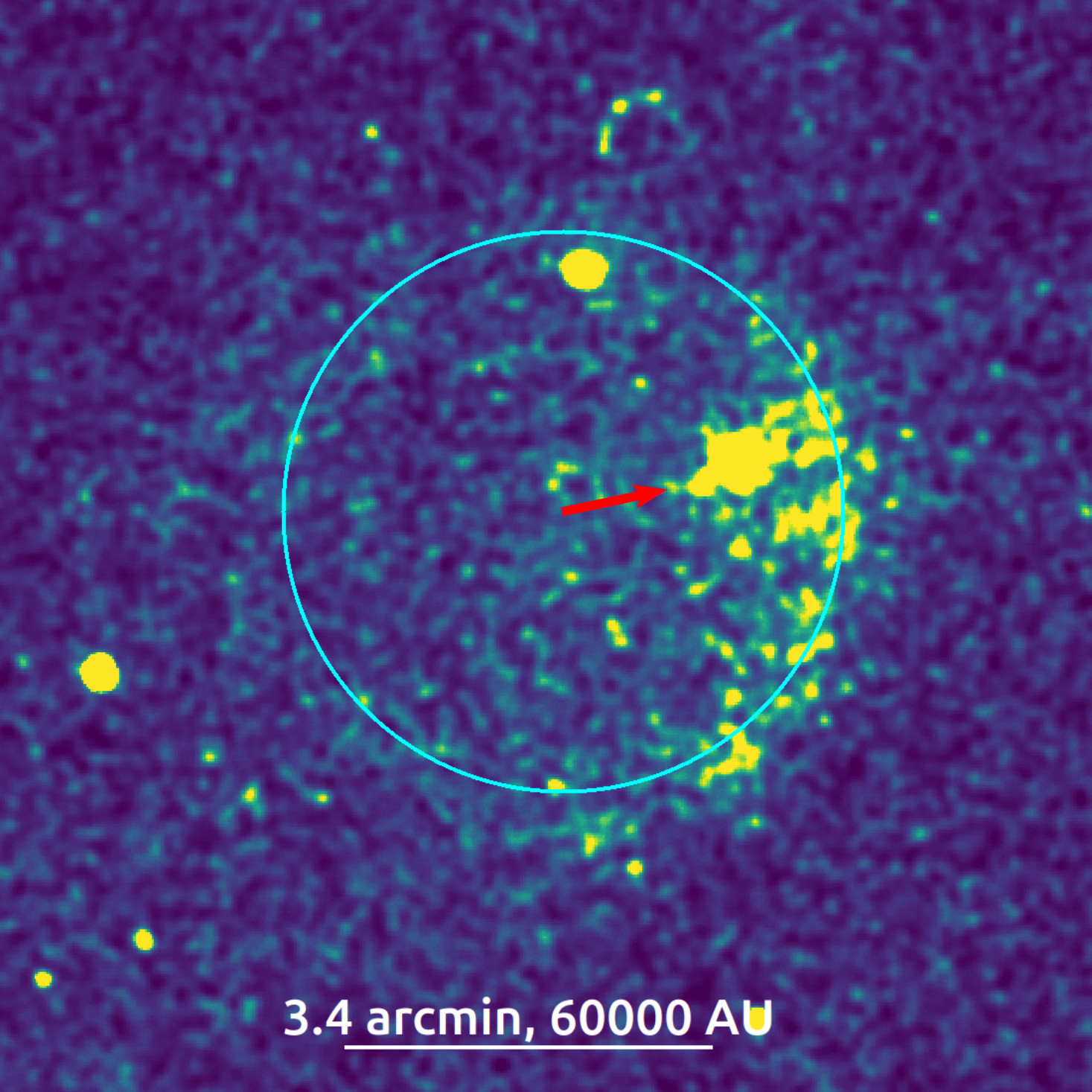}
        \caption{Galex FUV image}\label{uauv2}
        \end{subfigure}
    \begin{subfigure}{.49\linewidth}
                \includegraphics[width = \linewidth]{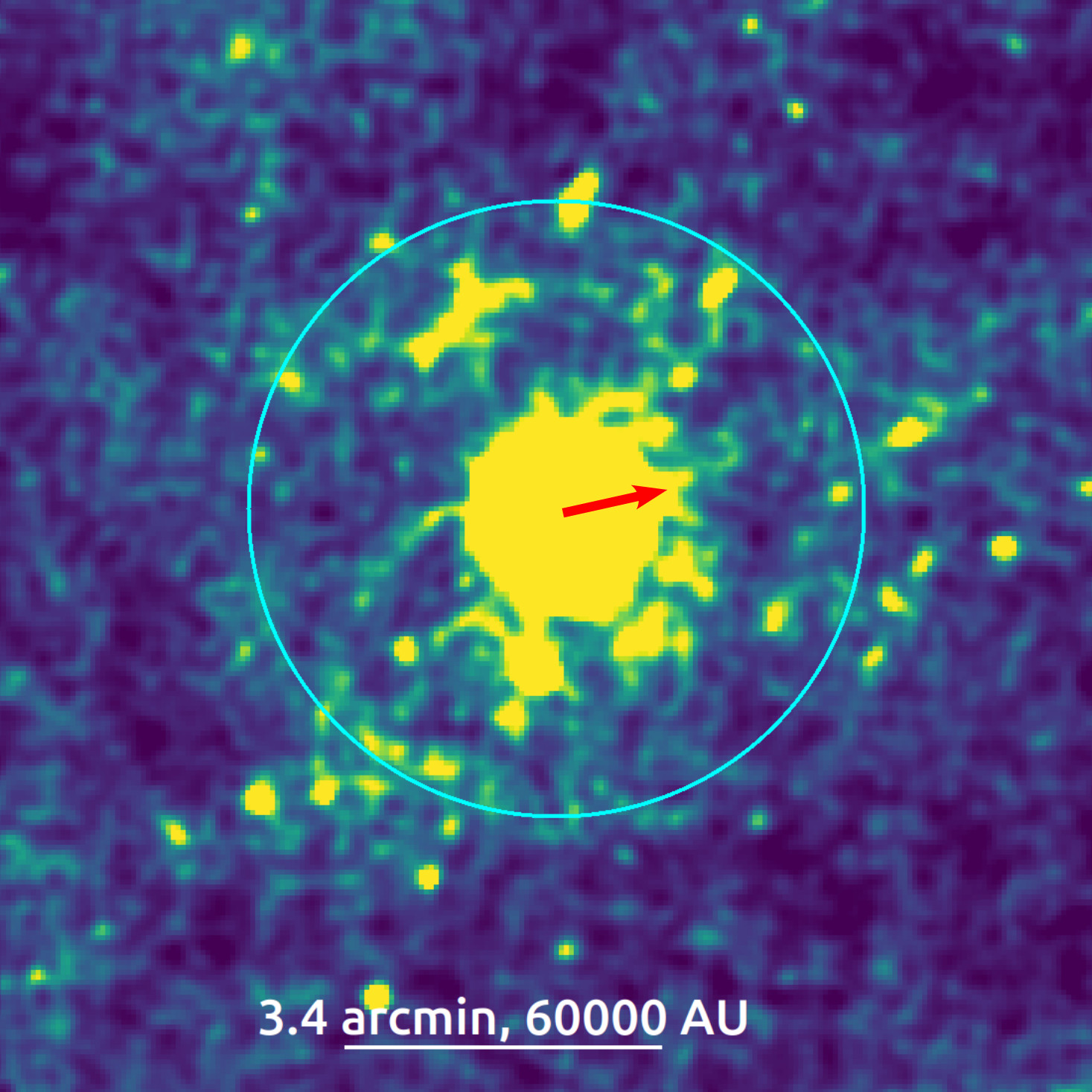}
        \caption{Herschel PACS160 image}\label{uair2}
        \end{subfigure}
        \caption{Extended emission seen around \object{U Ant} in FUV (left) and FIR (right, in units of Jy/pixel). While the central FIR emission present close to the star is not visible in the FUV, the larger arc located further away from the star is present in both images and is tentatively marked by the cyan dashed circle (radius of 3\arcmin ). The arrows are the same as in Fig.~\ref{fuvall}.} \label{uant}
\end{figure}

 The circumstellar environment of \object{U Ant} has been well studied. A first detection of an overall spherically symmetric detached CO shell around the star was reported by \citet{olofsson1988}. The CO shell emission, located at a distance of about 43\arcsec, has been imaged in detail using ALMA \citep{kerschbaum2017}. The studies of \citet{maercker2010} and \citet{kerschbaum2011} detected a dust shell counterpart to the CO shell in dust-scattered light and infrared dust emission (see Fig.~\ref{uair}), respectively. This detached shell most likely originates from a short period of high ML that the star underwent about 2700 years ago, when the MLR could have been as high as $\mathrm{10^{-5}\,M_{\sun}\,yr^{-1}}$. This shell is not visible in the UV range.

\par In a study based on IRAS 100 $\mu$m observations, \citet{izumiura1997} reported the detection of a larger dust shell with a radius of $\approx$ 3\arcmin, about four times larger than the inner CO shell. At low levels, the Herschel PACS 160\,$\mu$m image confirms the structure reported by \citet{izumiura1997} (Fig.~\ref{uair}), but the IR emission is quite diffuse so it is hard to tell what its full extent is. However, this outer shell is clearly seen in the FUV image (Fig.~\ref{uauv}) marked by the cyan circle that has a radius of 3\arcmin\ \citep[a similar size was deduced in a recently published independent study by][]{sahai2023}. With an increased intensity in the direction of motion of the star, the FUV emission resembles a bow shock and is located at a distance of about 50\,000 AU. Assuming an expansion velocity of $\mathrm{10\,km\,s^{-1}}$ the kinematic age of the shell would be $t_{\mathrm{kin}} \approx$ 23\,700 years. Considering the correlation between the direction of motion of the star, the space velocity ($\mathrm{42\,km\,s^{-1}}$) and the emission intensity in the FUV, this structure is most likely caused by wind--ISM interaction.

\subsection{R Doradus}
\indent\indent \object{R Dor} is a relatively evolved M-type SRb variable star located only 44 pc away, with a $T_{\mathrm{eff}}\mathrm{\approx 2700\,K}$ \citep{ohnaka2019} and a luminosity of about $\mathrm{2800\, L_{\sun}}$ \citep{ohnaka2019}. Two periods have been reported for this object (172 or 338 days). \citet{bedding1998} suggested that the star oscillates between two different pulsation modes (first and third overtone). 

\par \object{R Dor}'s position in the IRAS colour--colour diagram (Fig.~\ref{vdvdiag}) suggests that it has a tenuous CSE. This star is rather well-studied in CO line emission, and its MLR has been reliably estimated to be of the order of $\mathrm{ 10^{-7}\,M_{\sun}\,yr^{-1}}$ \citep{maercker2016, miora2021} (our estimate for its present-day MLR, see Table~\ref{params}, is in agreement with this value). High-angular-resolution molecular line studies suggest that \object{R Dor} has a low-mass close-in companion, and that this may have an effect on the structure of its CSE, at least on the inner parts \citep{vlemmings2018}.

\par
\begin{figure}[htb]
\centering
        \begin{subfigure}{.49\linewidth}
                \includegraphics[width = \linewidth]{rdor_fuvang.jpeg}
                \caption{Galex FUV image} \label{rduv}
        \end{subfigure}
        \begin{subfigure}{.49\linewidth}
                \includegraphics[width = \linewidth]{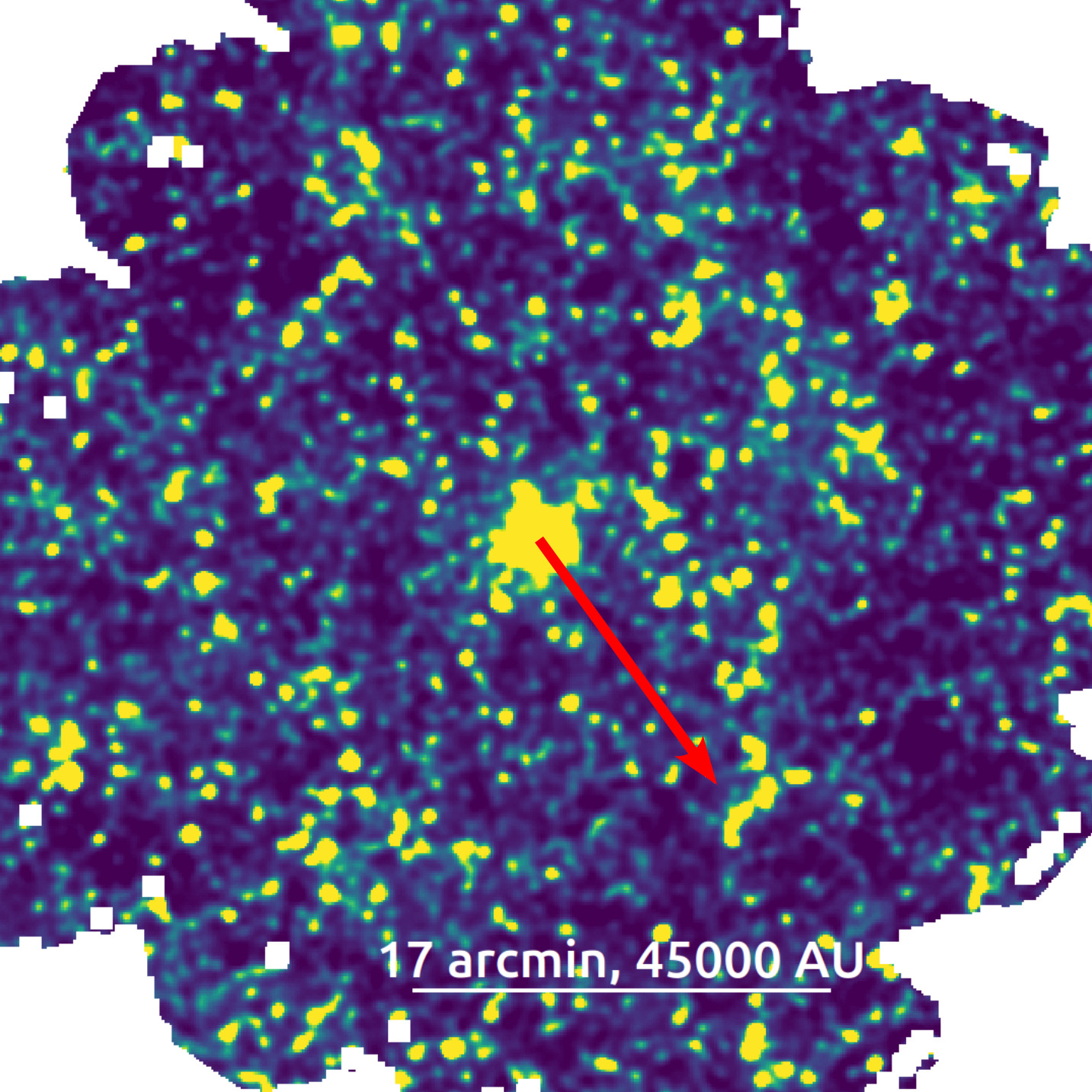}
                \caption{Herschel SPIRE250 image} \label{rdir}
        \end{subfigure}
    \begin{subfigure}{.49\linewidth}
                \includegraphics[width = \linewidth]{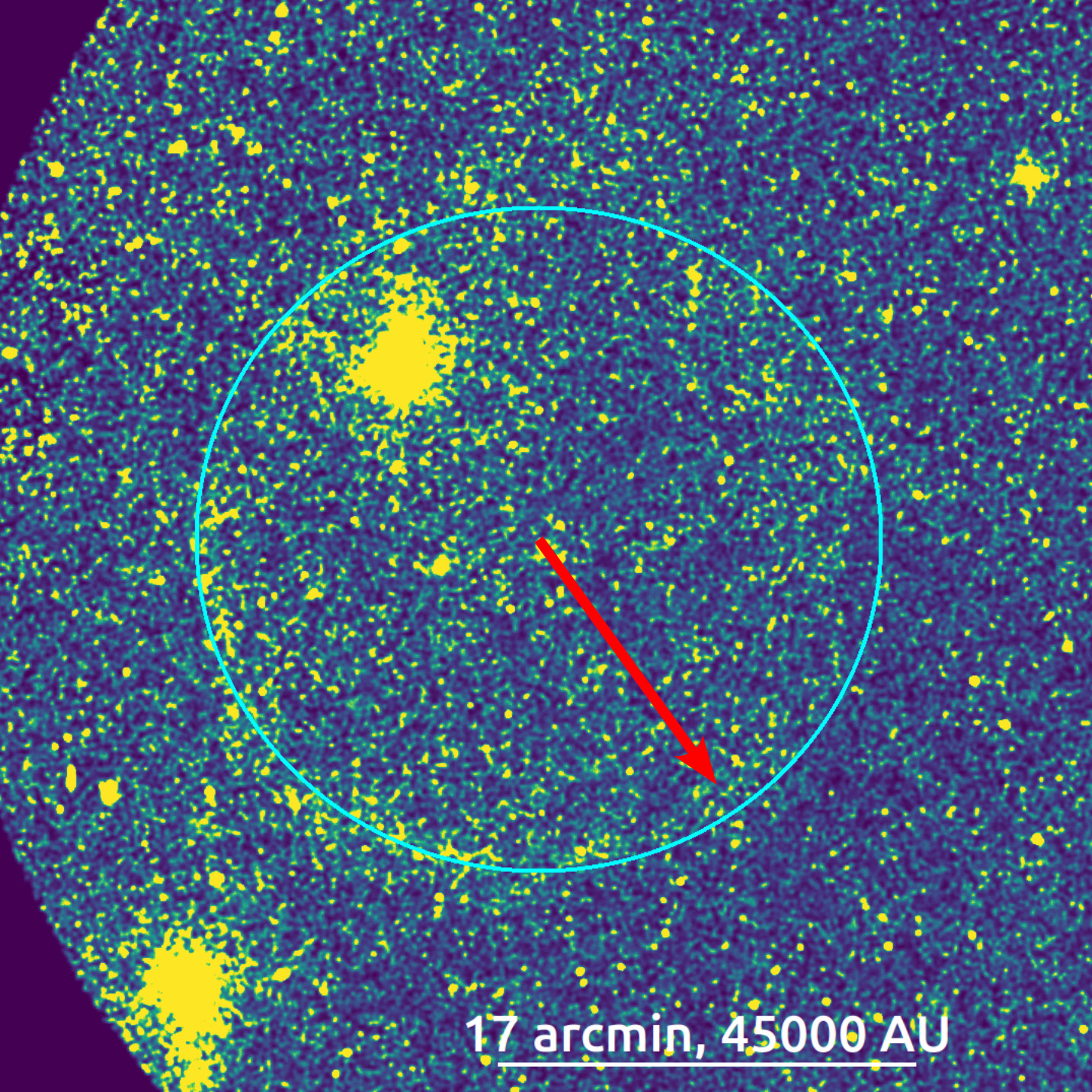}
        \caption{Galex FUV image} \label{rduv2}
        \end{subfigure}
        \begin{subfigure}{.49\linewidth}
                \includegraphics[width = \linewidth]{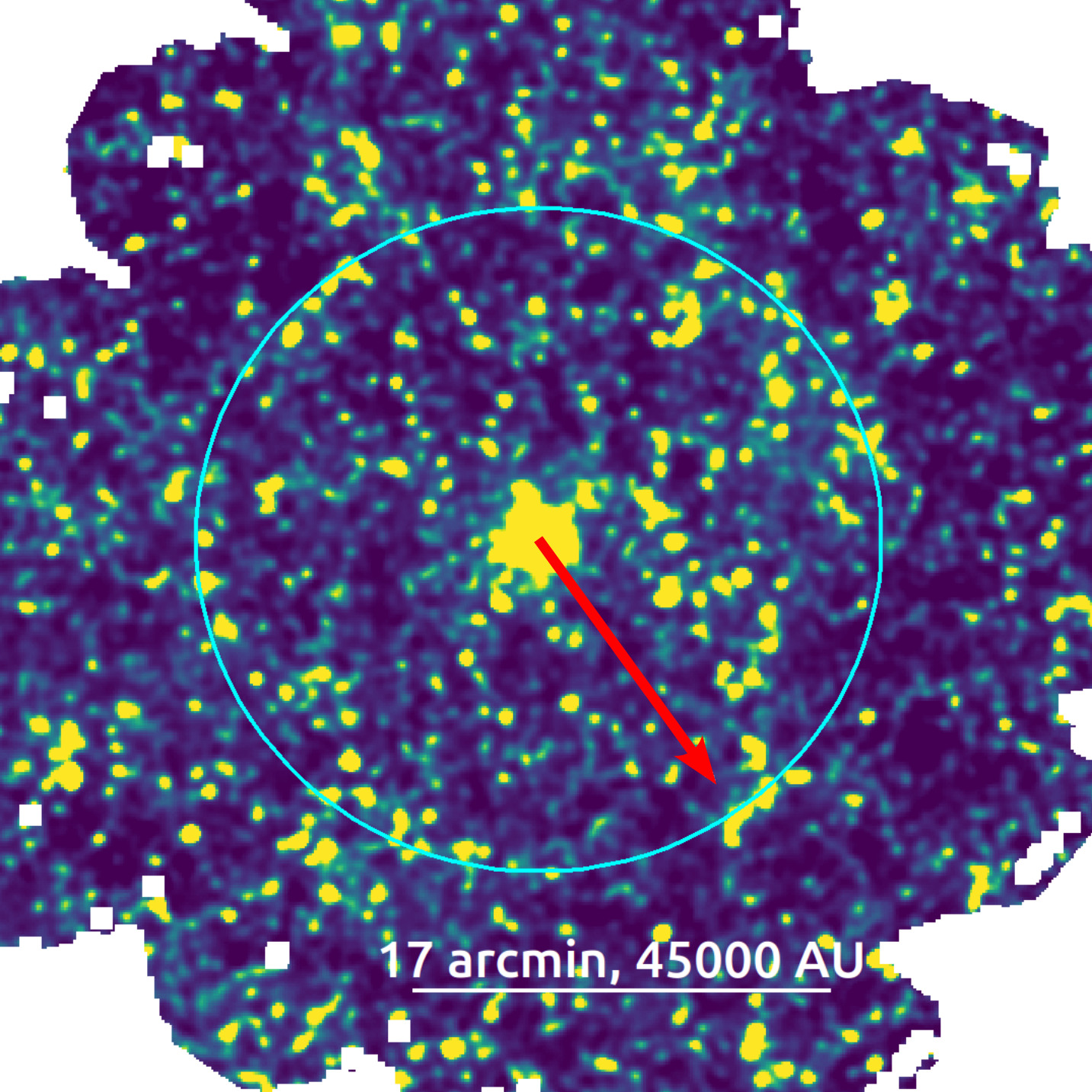}
        \caption{Herschel SPIRE250 image} \label{rdir2}
        \end{subfigure}
        \caption{Extended emission seen around \object{R Dor} in FUV (left) and IR (right). The cyan ellipse gives  the approximate location of the extended structure. The arrows are the same as in Fig.~\ref{fuvall}. The bright spot seen NE of R Dor in the Galex image is a hot subdwarf located about 1300\,pc away.} \label{rdor}
\end{figure}

 The emission seen in the FUV is largely circular and has no NUV counterpart. The same extended structure can be seen in the FIR (see Fig.~\ref{rdor}), although not as pronounced. With some tentative asymmetry towards the north in the FUV image (Fig.~\ref{rduv}), this ring structure has a radius of $R_{\mathrm{es}} \approx$ 37\,000 AU (13.5\arcmin, indicated by the dashed ellipse in Fig.~\ref{rdor}). This result is in agreement with an analysis recently published in parallel by \citet{ortiz2023}. Assuming a gas expansion velocity of $\mathrm{10\,km\,s^{-1}}$, the kinematic age of this structure would be $\approx$ 17\,500 years. Following its fairly circular shape and the lack of NUV emission, this extended structure may be caused by shocked $\mathrm{H_2}$ gas, be it wind--ISM or wind--wind interaction. The shock hypothesis could be supported by the fact that the space velocity of the star is a few times higher than the expansion velocity of its wind ($\mathrm{v_{s_{lsr}} = 40\,km\,s^{-1}}$ vs $\mathrm{10\,km\,s^{-1}}$). The seemingly stronger emission in other directions than that of the star's proper motion could be caused by $\mathrm{H_2}$ clumps (as postulated by \citealt{sanchez2015}), although the short exposure time is not sufficient to fully reveal whether this is the case or not. The clumpy aspect is slightly more evident in the FIR images (Fig.~\ref{rdir}).

\subsection{$\beta$ Gruis}
\indent\indent \object{$\beta$ Gru} is located only 54\,pc away (from Hipparcos parallax). Considered a SRb variable, its parameters gathered from the literature indicate an early-type AGB star, having an effective temperature $T_{\rm eff}$ = 3480\,K \citep{engelke2006}, a luminosity of $L = \mathrm{2500\,L_{\sun}}$ \citep{judge1986}, and a pulsation period $P = \mathrm{37\,days}$ \citep{otero2006}.

\par Using various UV emission lines (e.g. \ion{S}{I}, \ion{Si}{II}, \ion{C}{I}, \ion{Fe}{II}) originating in the outer atmosphere, the study of \citet{judge1986} set a lower limit of $\mathrm{5 \times 10^{-9}\, M_{\sun}\,yr^{-1}}$ for the MLR of the star. Using the obtained upper limit to the CO($J$\,=\,2–1) line and the MLR formula of \citet{ramstedt2008}, we obtain a crude upper limit to the gas-MLR of 10$^{-9}\,M_{\sun}\,{\rm yr}^{-1}$ (a gas expansion velocity of 7 km\,s$^{-1}$ and a circumstellar CO-to-H$_2$ abundance ratio of 4$\times$10$^{-4}$ were adopted). We note that the present-day ML that we computed via SED-fitting is consistent with this upper limit ($\mathrm{1.7\,\times\,10^{-9}\,M_{\sun}\,yr^{-1}}$). There is no indication of CO line emission from any detached shell.

\par Without substantial mass-loss, or with little to no dust emission \citep{sloan1998}, and based on its placement in the IRAS colour--colour diagram (Fig.~\ref{vdvdiag}), \object{$\beta$ Gru} appears to lack a CSE. However, recent radio observations show a potential departure from a black-body spectrum in the millimetre regime \citep{murphy2010, everett2020}. The presence of the essentially circular UV shell centred  on the star (Fig.~\ref{fuvallc}) indicate that \object{$\beta$ Gru}  underwent a significant episode of ML earlier in its evolution. All of this, combined with the lack of NUV or IR emission, suggests the shell seen in the FUV images could be attributed to wind--ISM interaction. We note that there is a lack of correlation between the strength of the FUV emission and the space motion of the star, but the low exposure time of the image is not enough to reveal more than that (i.e. whether there are any clumps of emission). With a radius of $R_{\mathrm{es}} \mathrm{\approx 58\,000\,AU}$ of the UV shell, its kinematic age is $t_{\mathrm{kin}} \approx$ 27\,500 years, assuming an expansion velocity of $\mathrm{10\,km\,s^{-1}}$. Alternatively, the object is still on the RGB and the UV shell is an effect of RGB wind--ISM interaction.

\subsection{W Hydrae}
\indent\indent \object{W Hya} is one of the most evolved stars in our sample. It is an M-type SRa variable located about 87 pc away \citep{miora2022}, and has an effective temperature of about 2600 K \citep{massalkhi2020}, a luminosity of $\mathrm{4730\,L_{\sun}}$ \citep[scaled for the distance given by][]{miora2022}, and a pulsation period of 390 days \citep{vogt2016}.

\par The IRAS  colour--colour diagram placement (Fig.~\ref{vdvdiag}) indicates that \object{W Hya} has no CSE, or a very thin CSE, but based on more recent studies and images there is a great deal of evidence that this is not true. The star has been studied extensively in the CO regime, and its present-day MLR is of the order of $\mathrm{10^{-7}\,M_{\sun}\,yr^{-1}}$ (\citealt{khouri2014}, \citealt{ramstedt2020}). The studies of \citet{muller2008} and \citet{vlemmings2011} have also detected the potential presence of a slow bipolar outflow, following HCN($J$\,=\,3--2) and SO($J$\,=\,5--4) observations. In a very recent study, combining ALMA observations of $\mathrm{^{12}CO}$($J$\,=\,2--1) and $\mathrm{^{29}SiO}$($J$\,=\,8--7) line emission and VLT visual and infrared observations, \citet{hoainhung2022} postulate a two-component scenario for the CSE of \object{W Hya}. They found an approximately spherical shell, at about 30 AU from the star, that is stable on a timescale of hundreds of years and that has an expansion velocity of 5 $\mathrm{km\,s^{-1}}$. They also present evidence for an elongated structure (10${\degr}$ west of north) which shows features characteristic of recent mass ejection; this component is found to be variable on a shorter timescale, years maybe even months. An older study by \citet{hawkins1990} suggested that \object{W Hya} might have previously experienced an episode of increased ML; this hypothesis is based on the fact that its present-day ML (measured at the time) was not high enough to justify the detection of an extended dust shell via IRAS observations.

\par
\begin{figure}[htb]
\centering
        \begin{subfigure}{.49\linewidth}
                \includegraphics[width = \linewidth]{whya_fuvang.jpeg}
                \caption{Galex FUV image} \label{whuv}
        \end{subfigure}
        \begin{subfigure}{.49\linewidth}
                \includegraphics[width = \linewidth]{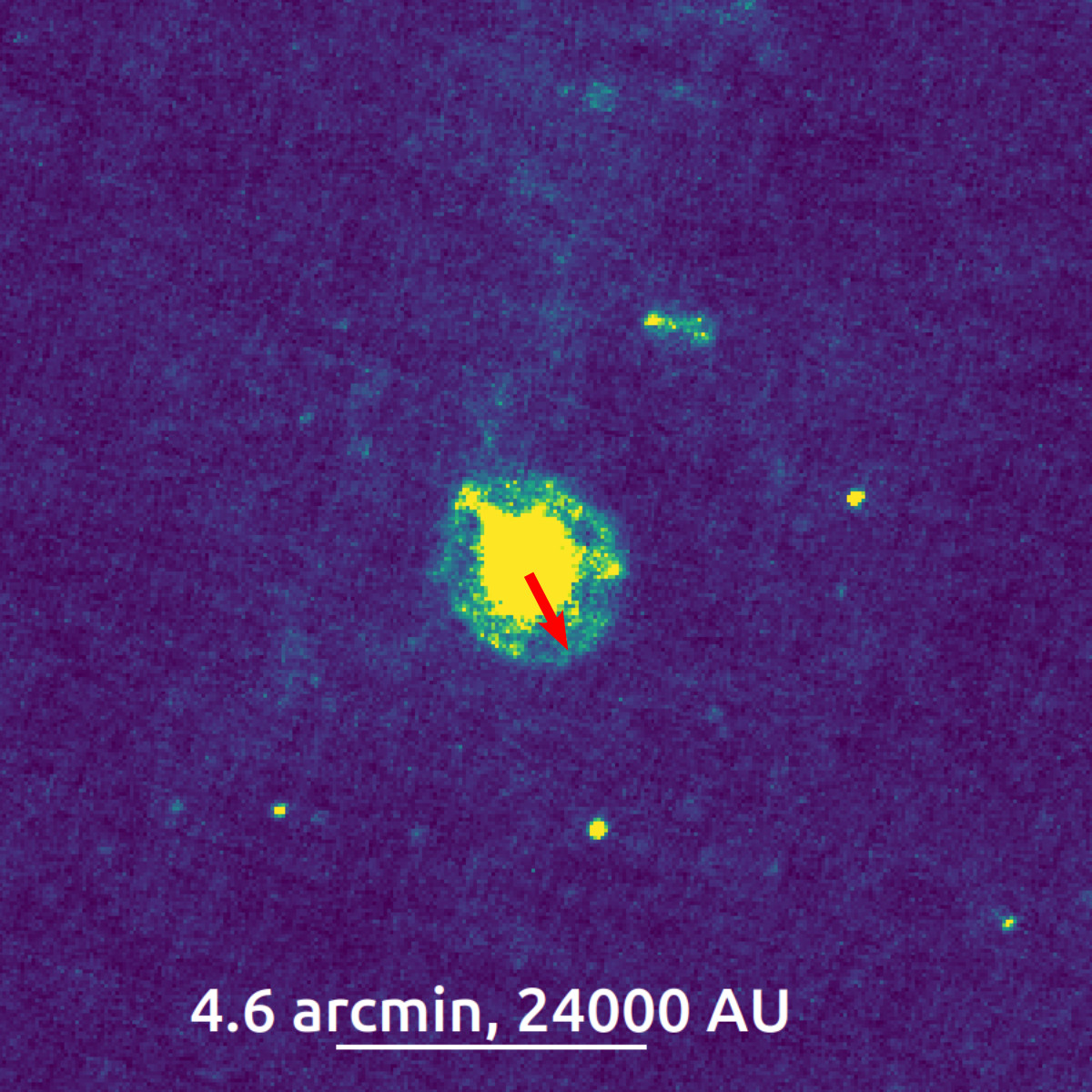}
                \caption{Herschel PACS160 image} \label{whir}
        \end{subfigure}
    \begin{subfigure}{.49\linewidth}
                \includegraphics[width = \linewidth]{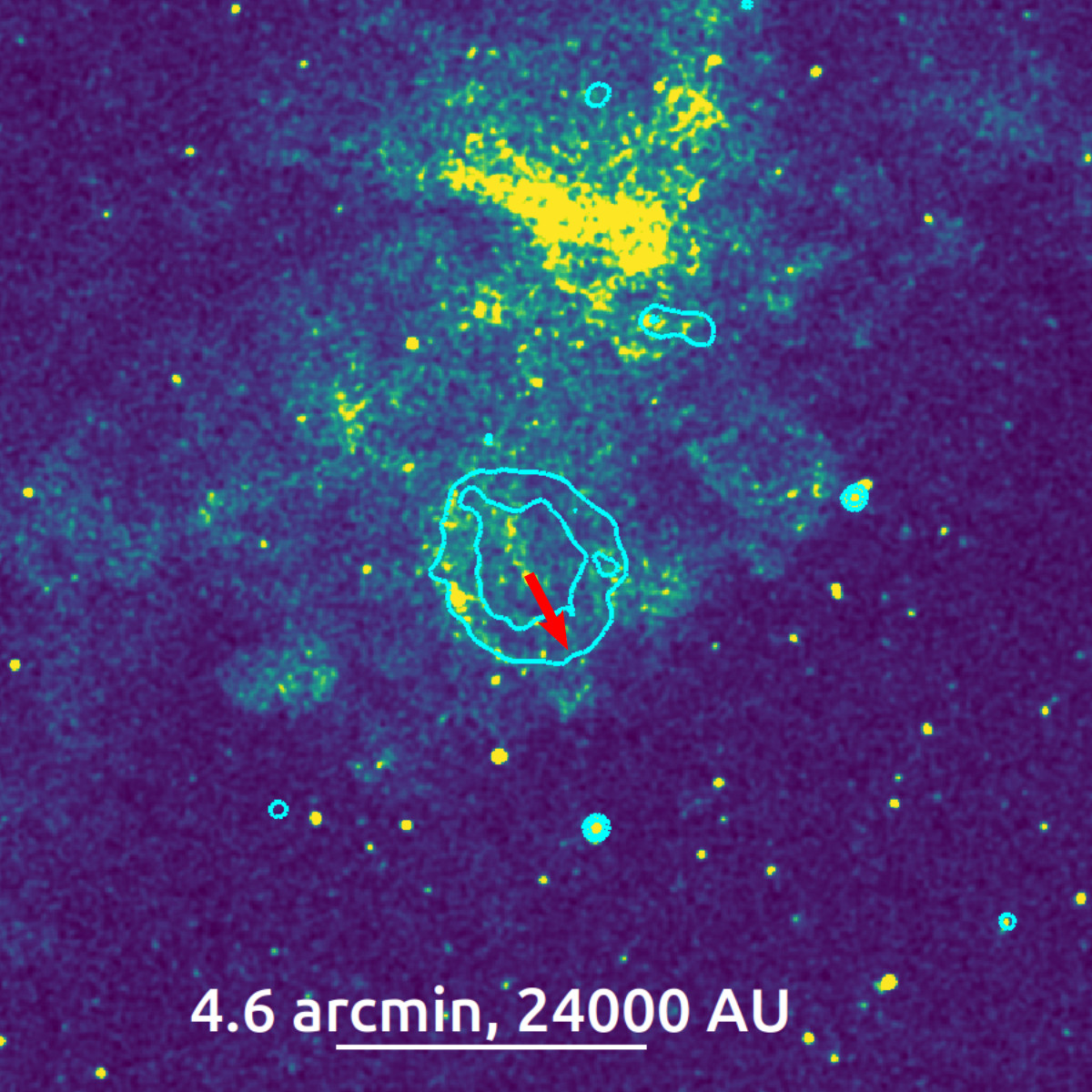}
        \caption{Galex FUV image} \label{whuv2}
        \end{subfigure}
        \begin{subfigure}{.49\linewidth}
                \includegraphics[width = \linewidth]{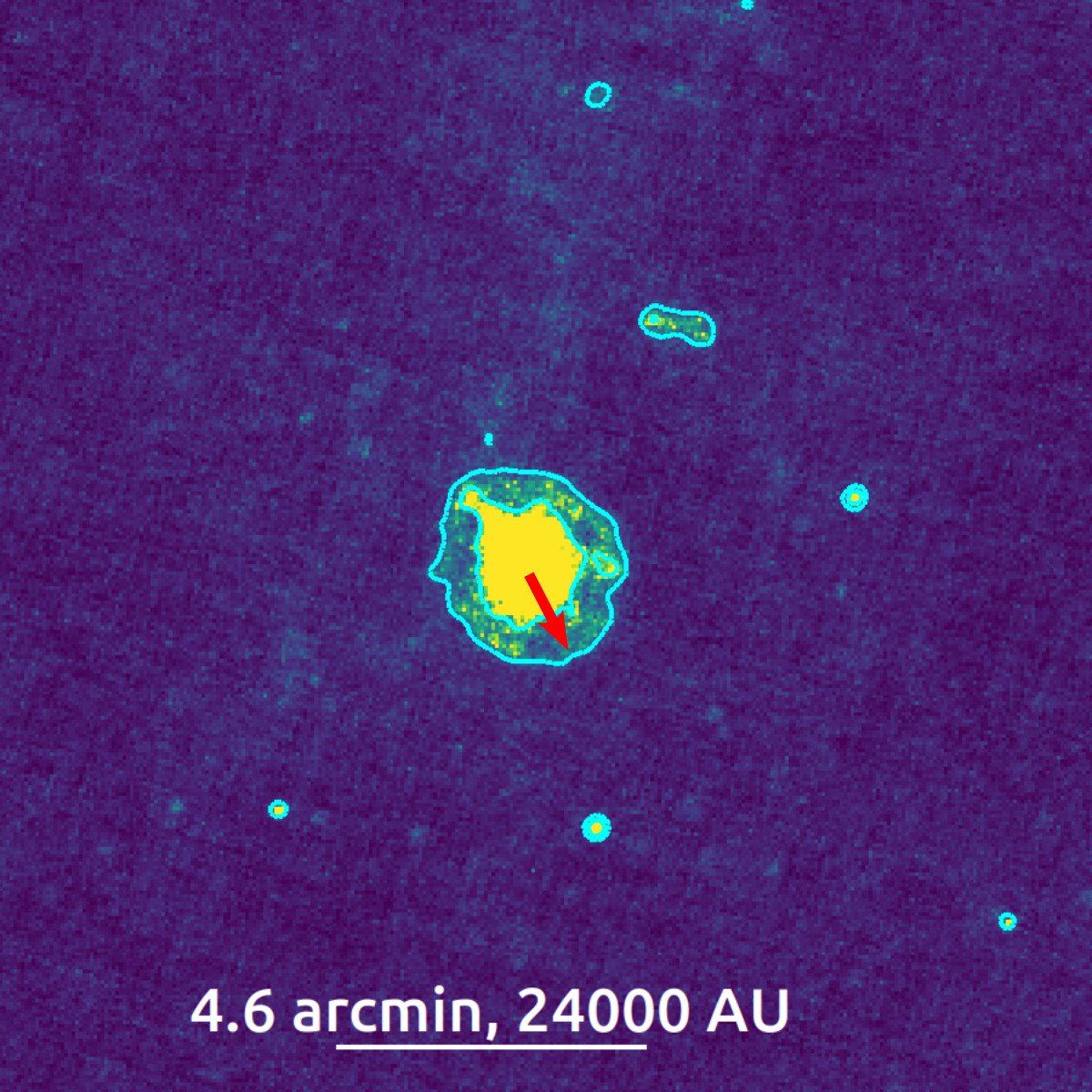}
        \caption{Herschel PACS160 image} \label{whir2}
        \end{subfigure}
        \caption{Extended emission seen around \object{W Hya} in FUV (left) and FIR. The emission seen in the IR is cospatial with the arc-like feature seen around the star in the FUV image, marked by the cyan contours (3 and 6\,$\sigma$ levels above the background) obtained from the infrared data. The arrows are the same as in Fig.~\ref{fuvall}.} \label{whya}
\end{figure}

 The GALEX FUV image shows extended emission that seems to consist of two components. A fairly circular shell is found closer to the star and is cospatial with FIR emission detected by the Herschel Space Observatory (see Fig.~\ref{whya}), a result reported in a recent study by \citet{sahai2023}. With a radius of $R_{\mathrm{es}}$ = 6\,250 AU ($\approx 1\arcmin$) and an assumed gas expansion velocity of $\mathrm{10\,km\,s^{-1}}$, the kinematic age of this shell corresponds to $\approx$ 3\,000 years. No NUV emission is seen around \object{W Hya} and once again the FUV component appears weaker in the direction of the star's proper motion. Considering its relative proximity to the star this structure could be caused by wind--wind interactions. The bright spot visible in the eastern part of the structure is most likely caused by the star \object{HD 120305}, located 130 pc away \citep{gaiadr3}. The second and more ambiguous component consists of very fuzzy emission seen to the north-west with respect to the star (Fig.~\ref{whuv},~\ref{whir}). This emission is very pronounced in the FUV, but it is also visible in the FIR images where it is more reminiscent of a tail-like structure. There is no obvious correlation between the direction of motion of the star and the direction in which the irregular emission is present. It is also hard to tell if the variable elongated structure found by \citet{hoainhung2022} north-west of the star could be in any way connected to this irregular FUV emission that extends to a distance beyond 24\,000 AU from \object{W Hya}.

\subsection{TV Piscium}
\indent\indent \object{TV Psc} is an M-type SR variable located about 190 pc away, with a pulsation period of only 49 days \citep{gcvs51}. Since it is not a well-studied object, literature data were not available regarding its luminosity, effective temperature, or MLR. Based on the effective temperature and luminosity we obtained by using the radiative transfer code \textit{More of DUSTY} \citep{groenewegendusty}, it falls in the same early-type AGB category as $\beta$ Gru (having a similar $T_{\mathrm{eff}}$ and $L$). The present-day MLR we computed, $\mathrm{1.5\times10^{-8}\,M_{\sun}\,yr^{-1}}$, also puts it at the low end of the AGB ML characteristics.

\par
\begin{figure}[htb]
\centering
        \begin{subfigure}{.49\linewidth}
                \includegraphics[width = \linewidth]{tvpsc_fuvang.jpeg}
                \caption{Galex FUV image} \label{tvfuv}
        \end{subfigure}
        \begin{subfigure}{.49\linewidth}
                \includegraphics[width = \linewidth]{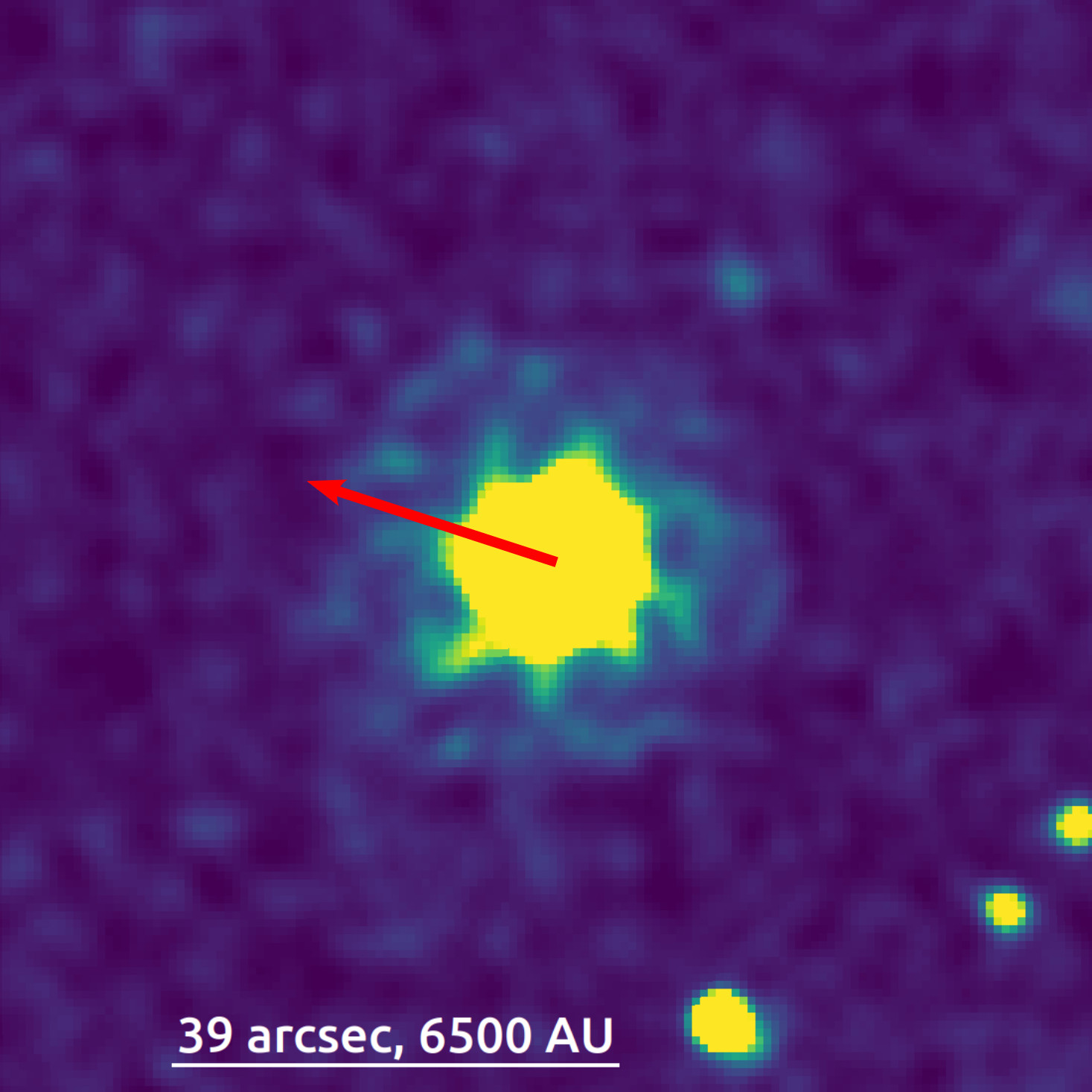}
                \caption{Galex NUV image} \label{tvnuv}
        \end{subfigure}
    \begin{subfigure}{.49\linewidth}
                \includegraphics[width = \linewidth]{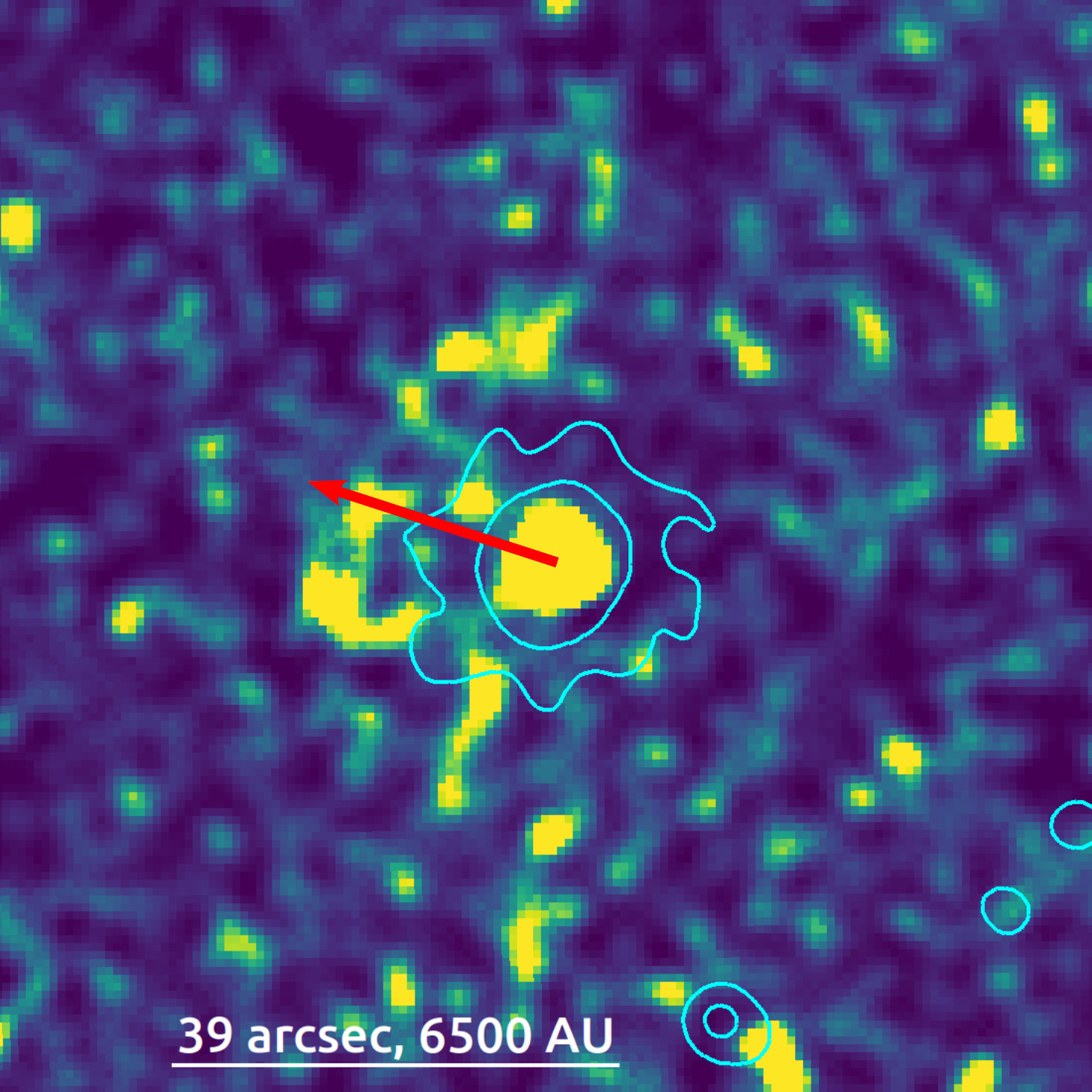}
        \caption{Galex FUV image} \label{tvfuv2}
        \end{subfigure}
        \begin{subfigure}{.49\linewidth}
                \includegraphics[width = \linewidth]{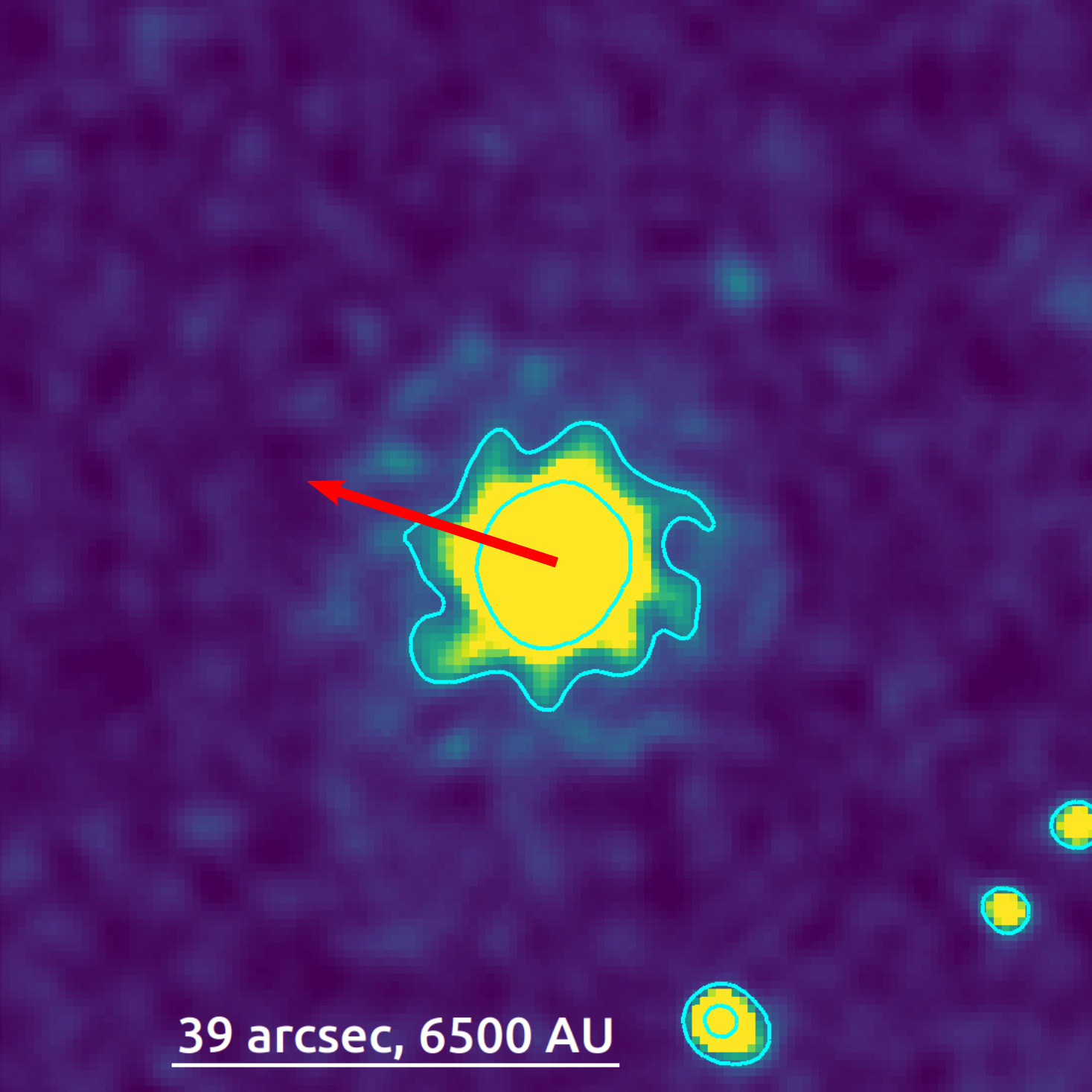}
        \caption{Galex NUV image} \label{tvnuv2}
        \end{subfigure}
        \caption{Extended emission seen around \object{TV Psc} in FUV (left) and NUV (right). The cyan   contours represent the 1 and 3\,$\sigma$ levels of emission above the background in the NUV data. The arrows are the same as in Fig.~\ref{fuvall}.} \label{tvpsc}
\end{figure}

 Despite its low MLR, indications of a bow-shock structure can be seen in the FUV image of \object{TV Psc} (Fig.~\ref{tvfuv}). No IR emission is present, but the NUV emission is rather strong. It is most likely associated with the star itself, being seen as a fairly pronounced circular shape around it (Fig.~\ref{tvnuv}). Although it is the smallest extended structure presented in this paper (located only $\approx$ 6600 AU from the star and corresponding to a kinematic age of 3100 years, assuming a gas expansion velocity of $\mathrm{10\,km\,s^{-1}}$), the position of FUV emission correlates well with \object{TV Psc}'s direction of motion. This correlation and the shape of the structure point to this being an effect of wind--ISM interaction, supported by the fact that the space velocity we computed (78 $\mathrm{km\,s^{-1}}$) is almost an order of magnitude higher than the assumed outflow velocity (10 $\mathrm{km\,s^{-1}}$), indicating that shocks are indeed the most likely culprit for the FUV emission in this case. A fairly pronounced and extended halo is seen in the NUV (and partly in the FUV), suggesting some other mechanism might be at play here. Its extent is given by the 1$\sigma$ level shown in Fig.~\ref{tvnuv2}, after which it drops in intensity significantly. This could be fluorescence caused by a companion, but the available data are not sufficient for a clear conclusion.

\subsection{RZ Sagittarii}
\indent\indent The only S-type star in the sample, \object{RZ Sgr,} is a Mira-type variable and is located approximately 432 pc away. Parameters gathered from the literature suggest that it is in a relatively advanced evolutionary stage on the AGB branch. It has an effective temperature of $2400$ K \citep{schoier2013}, a luminosity of $\mathrm{5000 \, L_{\sun}}$ \citep[scaled for the distance computed by \citet{miora2022} and based on the value of][]{winters2003}, and a period of approximately 223 days \citep{gcvs51}.

\par \object{RZ Sgr} has the highest MLR in our sample, of the order of $\mathrm{10^{-6}\,M_{\sun}\,yr^{-1}}$ based on CO line data \citep{ramstedt2009}. Multiple IRAS studies report it as having an extended structure at 60 and 100 $\mu$m \citep{fouque1992, young1993II}, and \citet{whitelock1994} reported some optical nebulosity around the star that might be related to the extended FIR structure. More recently \citet{miora2021} observed the CSE of \object{RZ Sgr} in CO($J$\,=\,2--1) and ($J$\,=\,3--2) emission, using the Atacama Compact Array. They report an inner CO CSE, with a size of $\approx$ 6\,600 AU, which shows signs of anisotropies (mainly towards the north with respect to the star) and even a potential extended spiral structure.

\par
\begin{figure}[htb]
\centering
    \begin{subfigure}{.49\linewidth}
                \includegraphics[width = \linewidth]{rzsgr_fuvang.jpeg}
                \caption{Galex FUV image} \label{rzfuv}
        \end{subfigure}
    \begin{subfigure}{.49\linewidth}
                \includegraphics[width = \linewidth]{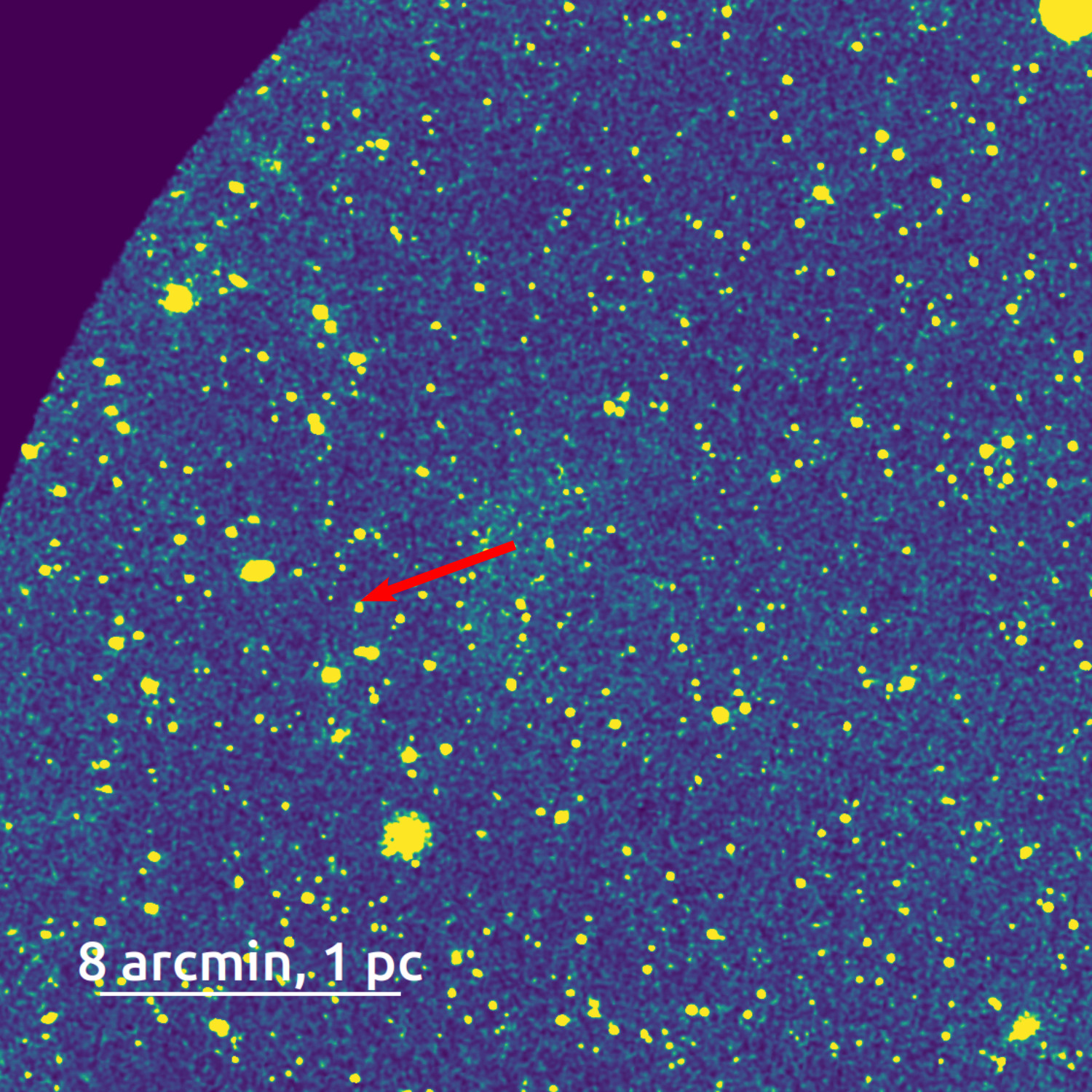}
                \caption{Galex NUV image} \label{rznuv}
        \end{subfigure}
        \begin{subfigure}{.49\linewidth}
                \includegraphics[width = \linewidth]{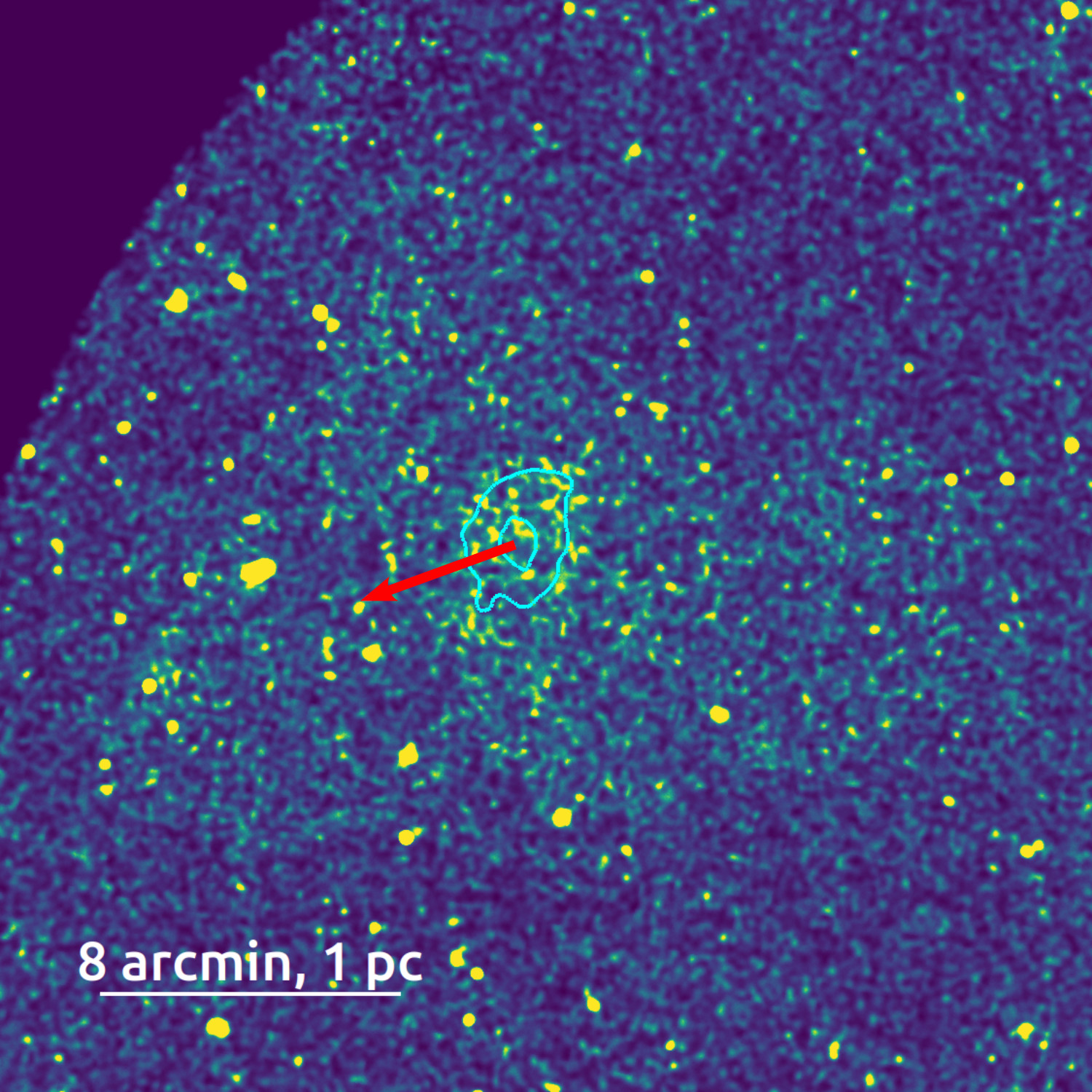}
                \caption{Galex FUV image} \label{rzfuv2}
        \end{subfigure}
        \begin{subfigure}{.49\linewidth}
                \includegraphics[width = \linewidth]{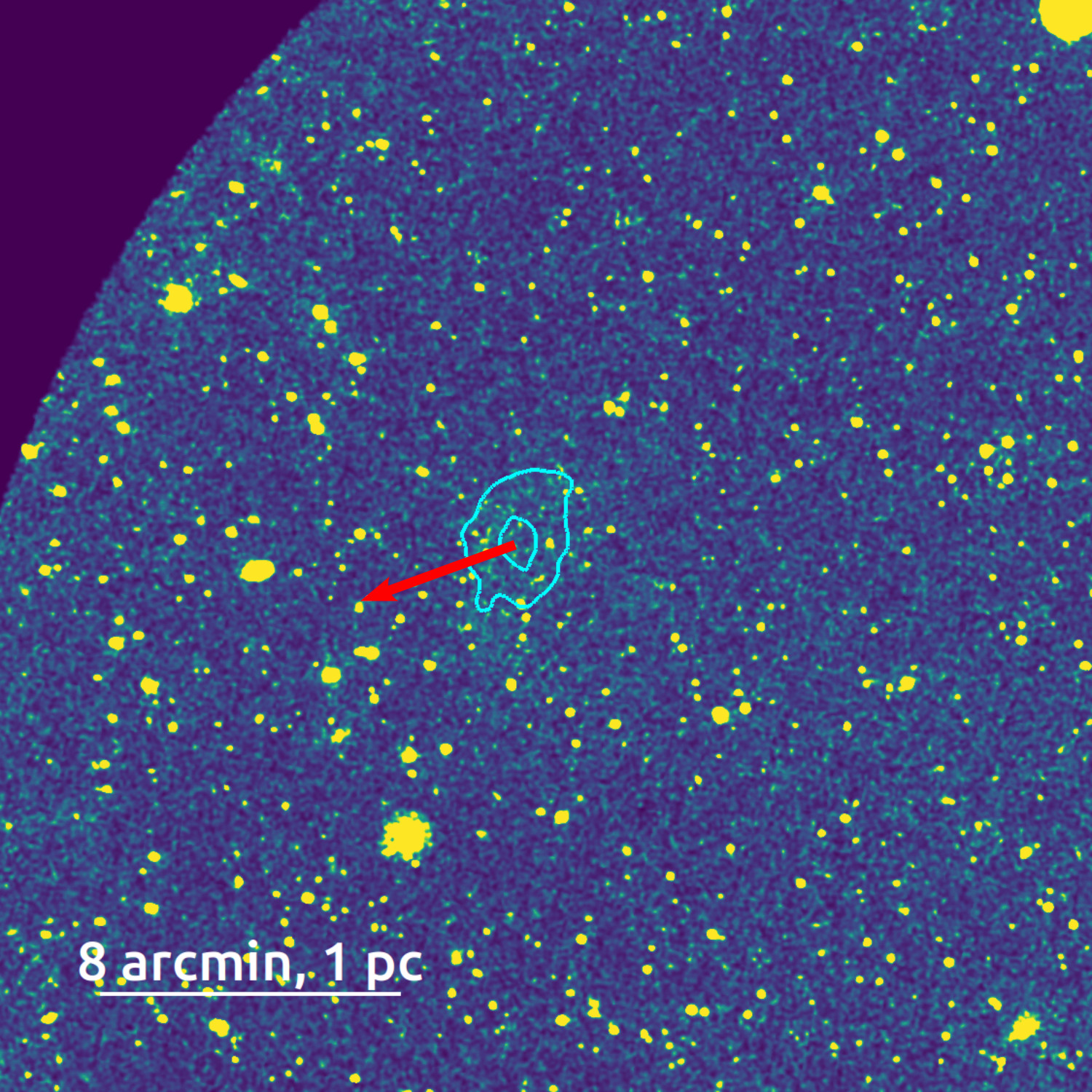}
                \caption{Galex NUV image} \label{rznuv2}
        \end{subfigure}
    \begin{subfigure}{.49\linewidth}
                \includegraphics[width = \linewidth]{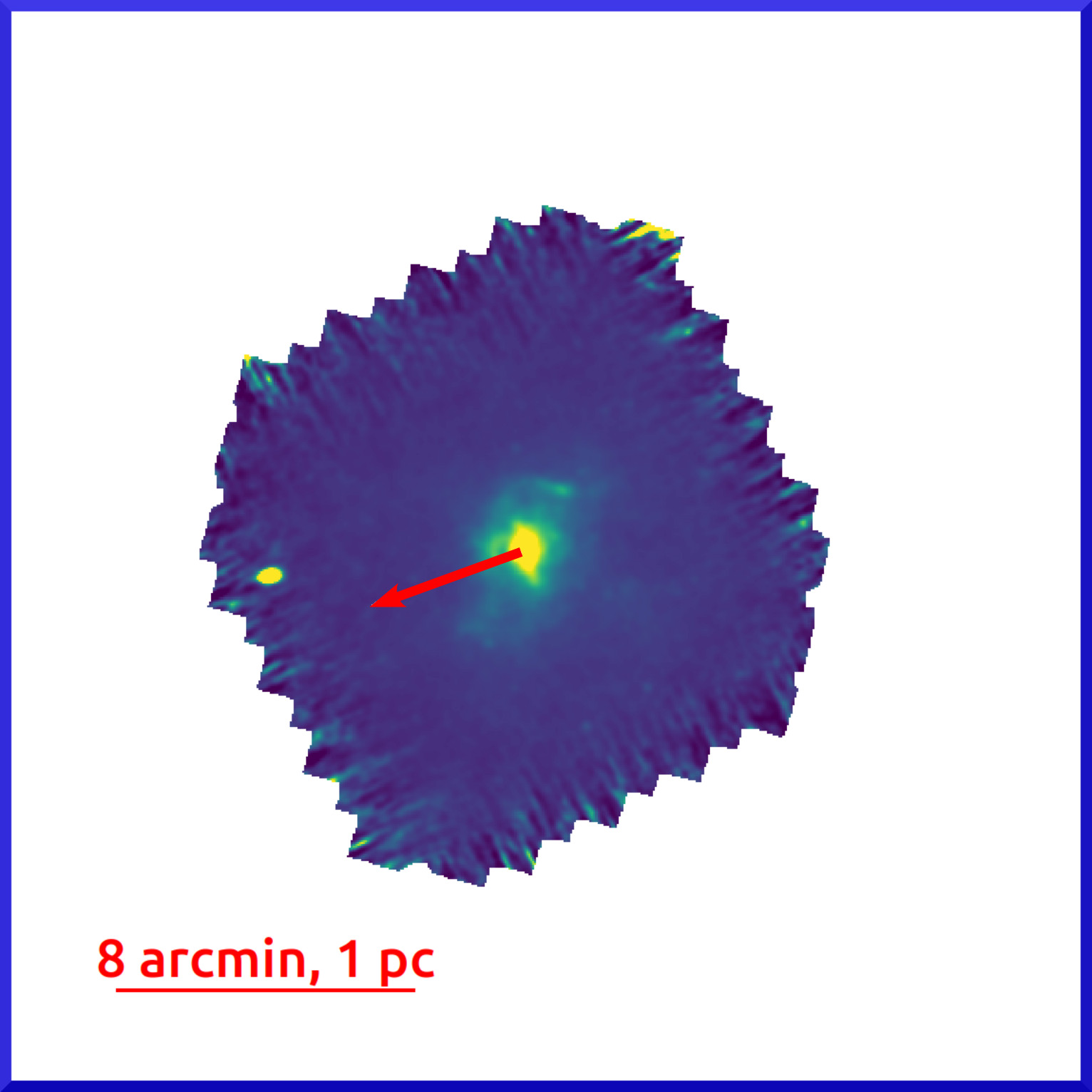}
                \caption{Herschel PACS160 image} \label{rzir}
        \end{subfigure}
        \begin{subfigure}{.49\linewidth}
                \includegraphics[width = \linewidth]{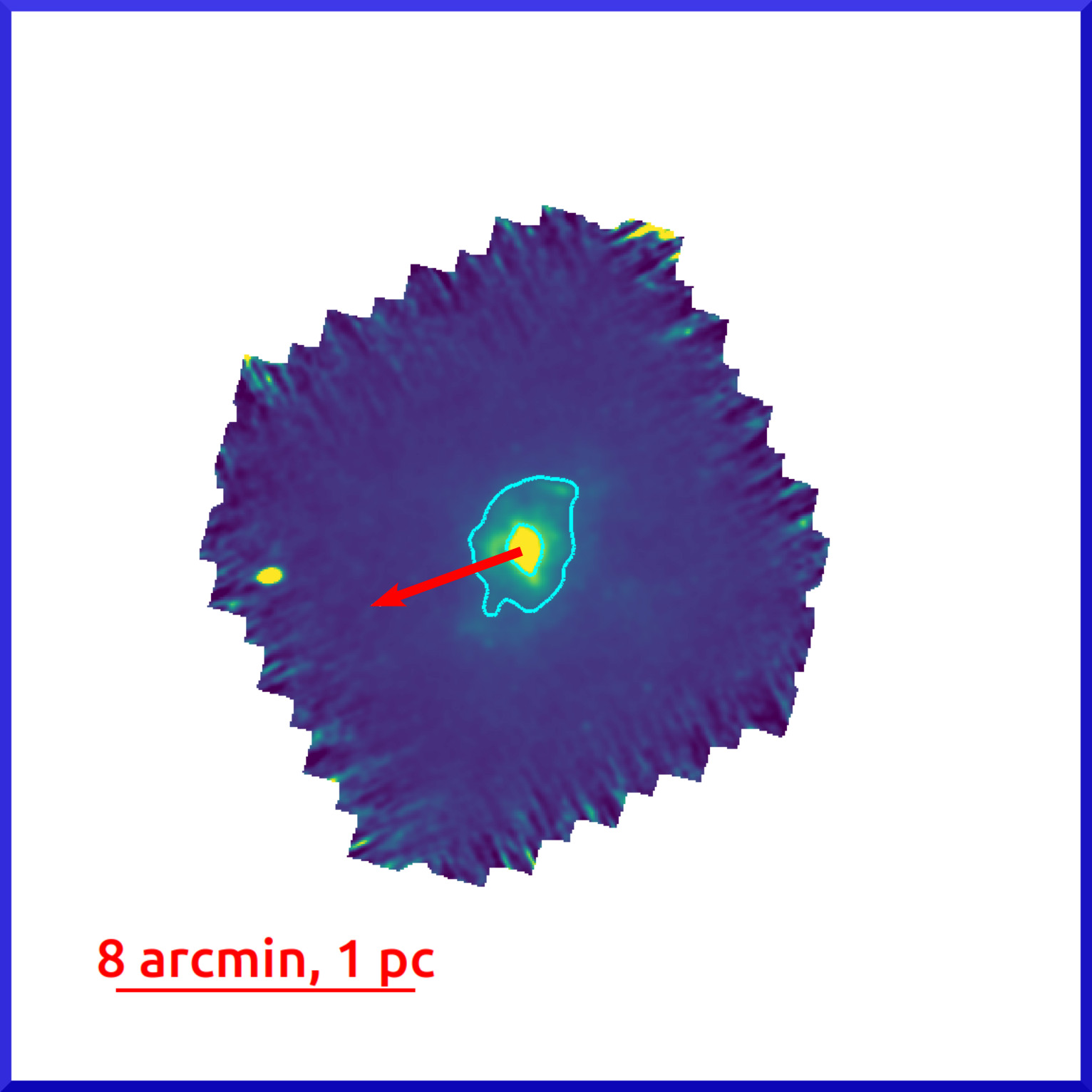}
                \caption{Herschel PACS160 image} \label{rzir2}
        \end{subfigure}
        \caption{Extended emission seen around the star \object{RZ Sgr} in FUV (panels a and c), NUV (panels b and d), and FIR (panels e and f). The emission seen in the bottom images is cospatial with the more pronounced central emission seen in the Galex images (FUV and NUV). The cyan contours represent the 2 and 5\,$\sigma$ levels of emission above the background in the FIR data. The arrows are the same as in Fig.~\ref{fuvall}.} \label{rzsgr}
\end{figure}

 Herschel FIR images of \object{RZ Sgr} show some extended emission surrounding the star (see Fig.~\ref{rzir}). The emission has a tentative eye shape, and a potential bipolar outflow can be seen. It is interesting that this FIR emission is cospatial with the central part of the FUV emission seen in Fig.~\ref{rzfuv} and also with the NUV emission (Fig.~\ref{rznuv}). A close inspection of Fig.~\ref{rzfuv} reveals that the FUV emission covers two different regions: a central part where the FUV emission is stronger and that is also cospatial with the FIR and NUV observations, and a fainter but much more extended ellipsoidal part with some potential asymmetry towards the north. The placement of this star in the IRAS colour--colour diagram (Fig.~\ref{vdvdiag}) would support this two-outflow structure as it lies in a region mainly occupied by objects having hot O-rich material separated from more distant cold dust. The studies of \citet{sahai1995} and \citet{jorissen1998} first postulated a two-outflow scenario (one of which is slower) following their CO observations, based on the fact that the expansion velocity derived from the CO($J$\,=\,2--1) line was not the same as the one derived from the CO($J$\,=\,1--0) line. 

\par Wind--wind or wind--ISM interactions could be the reason for the extended structure seen only in the FUV, considering the space velocity of the star of $\mathrm{41\,km\,s^{-1}}$. Using a constant gas expansion velocity of $\mathrm{10\,km\,s^{-1}}$, this structure has an outer radius of as much as $R_{\mathrm{es}} \approx$ 180\,000 AU and a kinematic age of 85\,000 years. Targeted observations would be necessary for gathering more information about this very large and faint structure. The origin of the diffuse inner region that is visible in all three wavelength ranges (FUV, NUV, FIR) is unclear.

\subsection{DM Tucanae}
\indent\indent \object{DM Tuc} is another M-type SRb variable. Its MLR, of the order of $\mathrm{10^{-7}\,M_{\sun}\,yr^{-1}}$ \citep{winters2003}, and other stellar parameters indicate it is in a more advanced stage of the AGB phase in comparison to  \object{$\beta$ Gru} or \object{TV Psc}. It is located at a distance of 232 pc and has a variability period of 75 or 145 days \citep{gcvs51}, which suggests the star is most likely oscillating between two different pulsation modes. 

\par
\begin{figure}[htb]
\centering
        \begin{subfigure}{.49\linewidth}
                \includegraphics[width = \linewidth]{dmtuc_fuvang.jpeg}
                \caption{Galex FUV image} \label{dmfuv}
        \end{subfigure}
        \begin{subfigure}{.49\linewidth}
                \includegraphics[width = \linewidth]{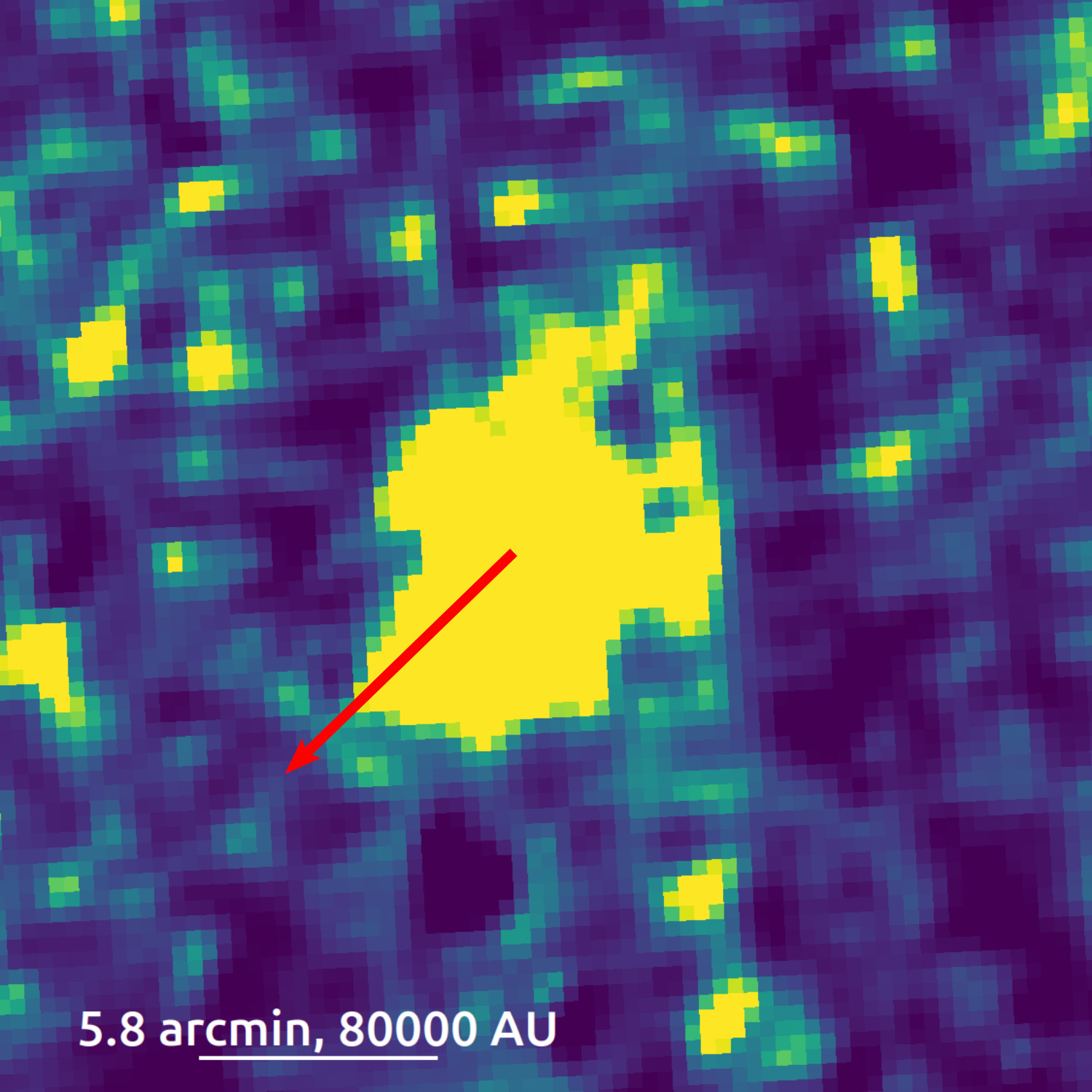}
                \caption{Akari WideS {\bfseries 90 $\mathrm{\mu m}$ image}} \label{dmfir}
        \end{subfigure}
        \begin{subfigure}{.49\linewidth}
                \includegraphics[width = \linewidth]{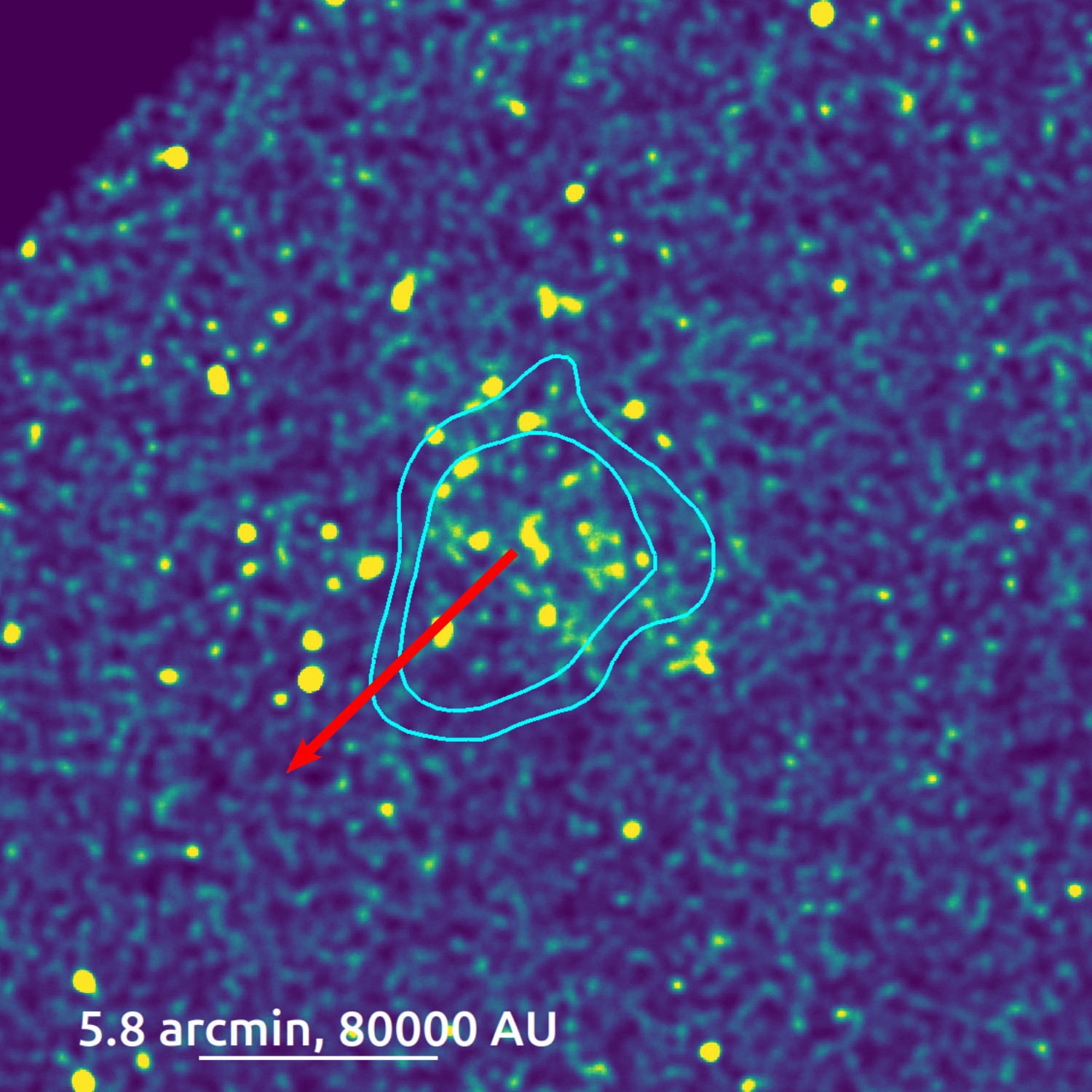}
        \end{subfigure}
    \begin{subfigure}{.49\linewidth}
                \includegraphics[width = \linewidth]{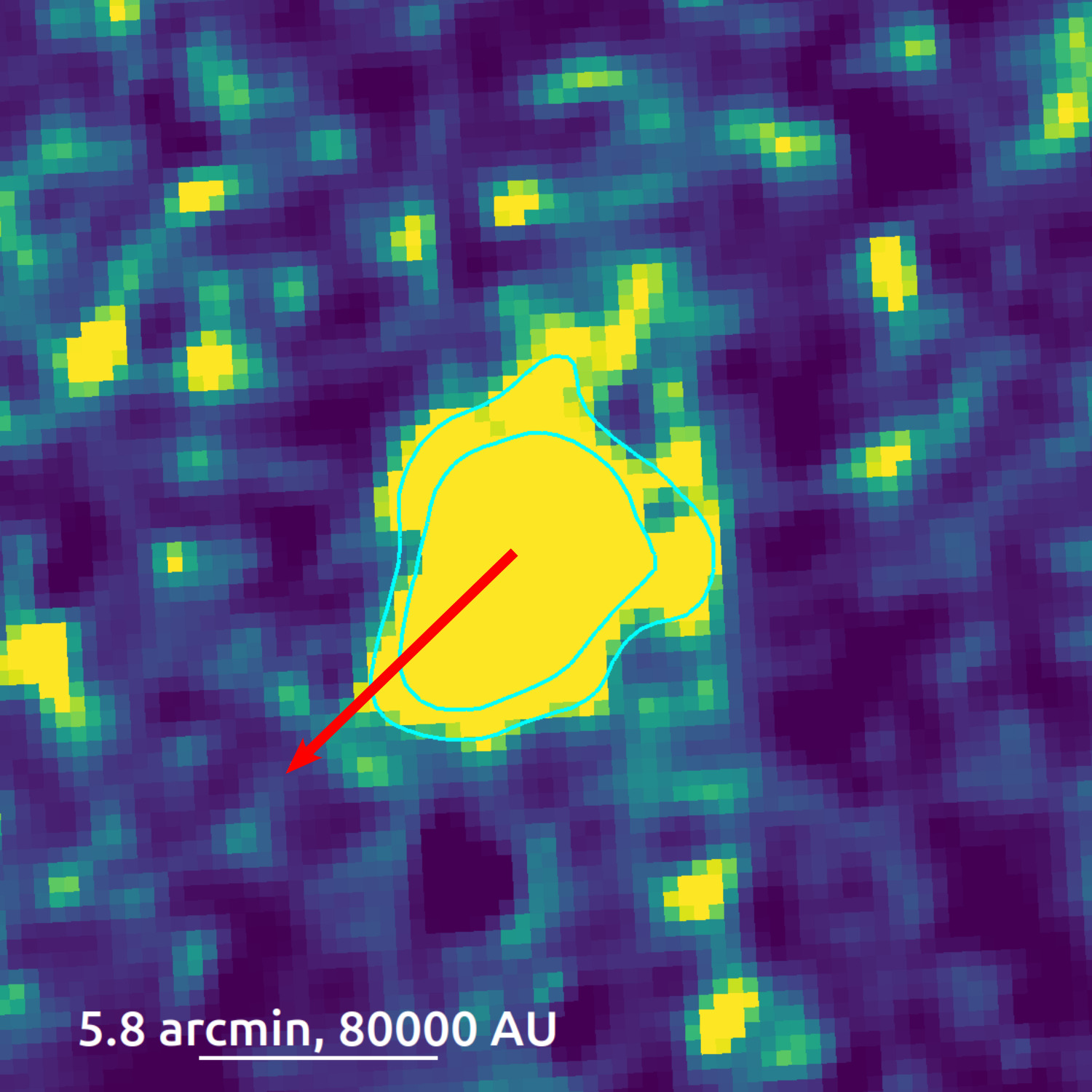}
        \end{subfigure}
        \caption{Extended emission seen around \object{DM Tuc} in FUV (left) and FIR (right). The cyan contours represent the 2 and 4\,$\sigma$ levels of emission above the background in the FIR data. The arrows are the same as in Fig.~\ref{fuvall}.}\label{dmtuc}
\end{figure}

 It is not a well-studied object, and little is known about its surroundings. The IRAS colour--colour diagram placement suggests \object{DM Tuc} could have a thin O-type CSE. Akari FIR observations support this (see Fig.~\ref{dmfir}); the $\mathrm{90\,\mu m}$ image shows some extended emission around the star. This emission is also partially cospatial with the irregular emission seen in the FUV observations (Fig.~\ref{dmfuv}). Figure~\ref{dmtuc} reveals that both the FUV and FIR irregular structures span a similar size, of the order of a few $10^4$ AU. What makes this case more interesting is the orientation of the emission. While the FUV emission seems to be oriented NE--SW, the FIR emission is oriented more along a NW--SE axis. Once more the FUV emission appears clumpy. The computed space velocity of the star is almost an order of magnitude higher than the assumed outflow velocity ($\mathrm{87\,km\,s^{-1}}$ vs 10 $\mathrm{km\,s^{-1}}$), suggesting that shocked $H_2$ could be a plausible source of emission.

\subsection{V420 Vulpeculae}
\indent\indent \object{V420 Vul} is a C-type Mira variable located about 2429 pc away. Apart from its variability period of 377 days \citep{gcvs51}, little is known about it. Using the radiative transfer code \textit{More of DUSTY} \citep{groenewegendusty} we calculated its effective temperature, luminosity, and present-day ML to be approximately $T_{\mathrm{eff}} \mathrm{= 3000\,K}$, $L \mathrm{ = 4000\, L_{\sun}}$, and $\dot{M}_{\mathrm{pd}} \mathrm{= 1.7\,\times\,10^{-7}\,M_{\sun}\,yr^{-1}}$, respectively.

\par Some extended emission is present around it in the FUV, but it is very faint and irregular. However, it is  in the same region where extended FIR emission can be observed (see Fig.~\ref{v420}). There is no obvious correlation between the proper motion of the star and the FUV emission. The computed space velocity of $\mathrm{132\,km\,s^{-1}}$ is more than an order of magnitude higher than the assumed 10 $\mathrm{km\,s^{-1}}$ outflow velocity. Combined with the apparent clumpiness of the FUV emission this suggests that the shocked molecular hydrogen scenario proposed by \citet{sanchez2015} could apply.

\begin{figure}[htb]
\centering
        \begin{subfigure}{.49\linewidth}
                \includegraphics[width = \linewidth]{v420_fuvang.jpeg}
                \caption{Galex FUV image} \label{v420fuv}
        \end{subfigure}
        \begin{subfigure}{.49\linewidth}
                \includegraphics[width = \linewidth]{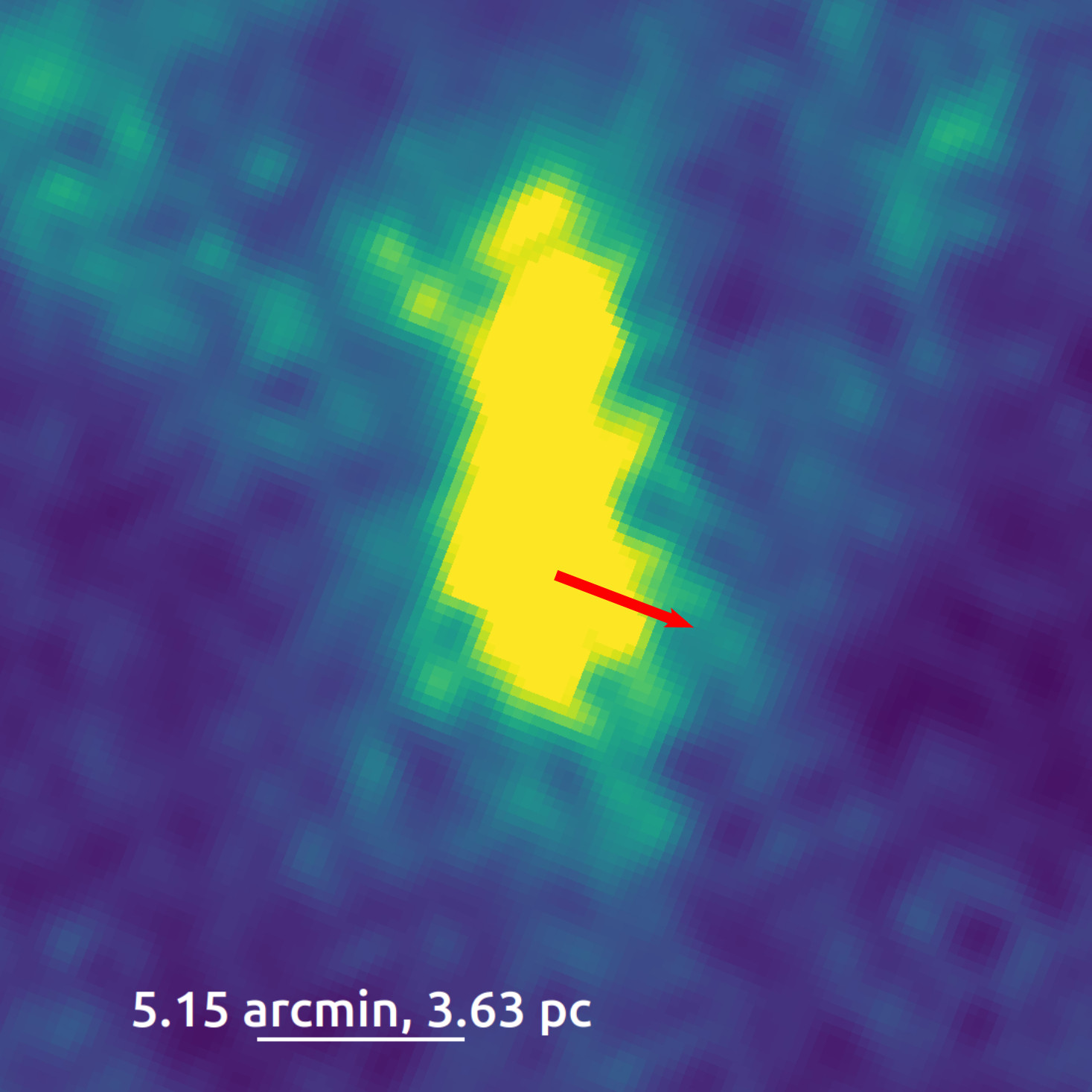}
                \caption{Akari WideS 90$\mathrm{\mu m}$ image} \label{v420fir}
        \end{subfigure}
    \begin{subfigure}{.49\linewidth}
                \includegraphics[width = \linewidth]{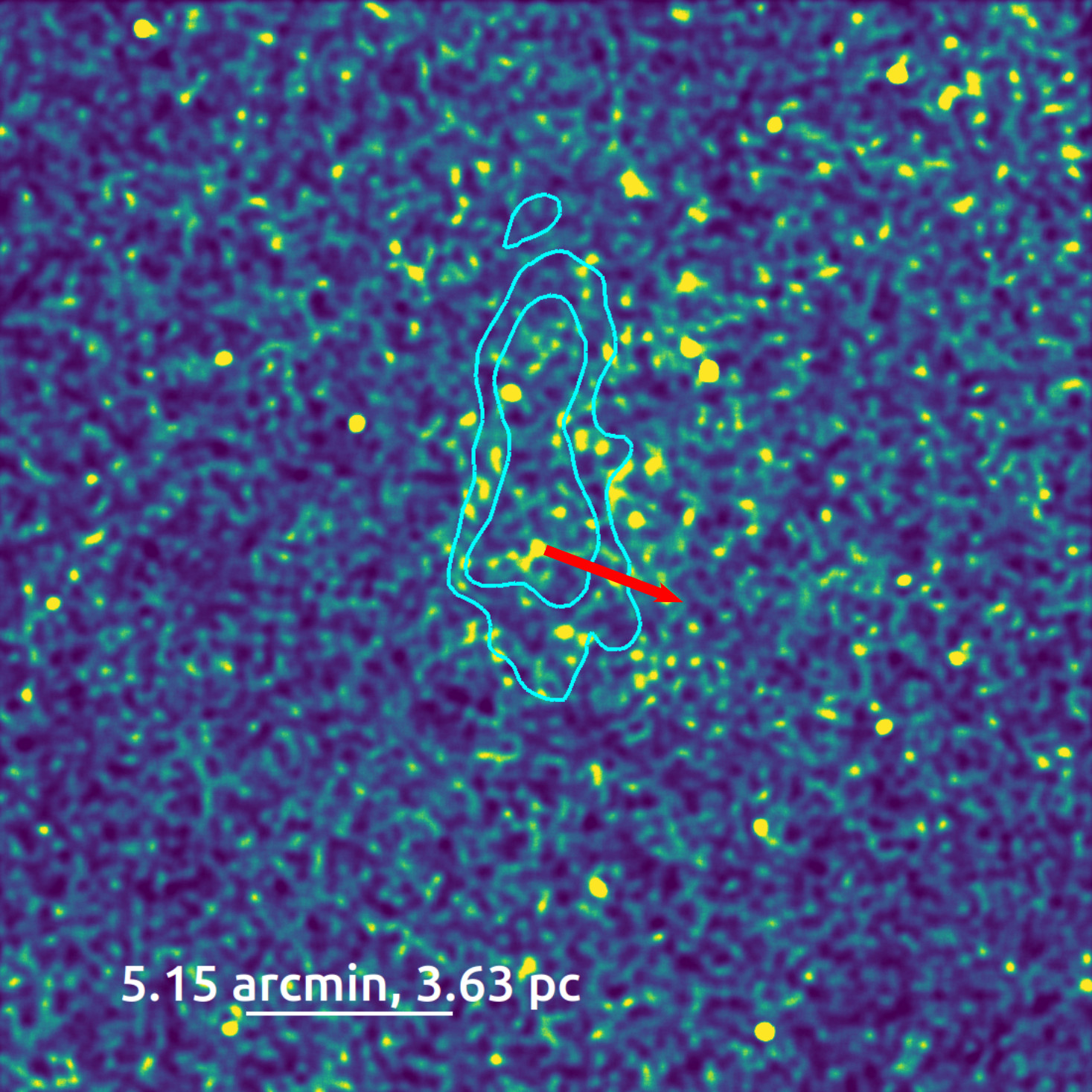}
        \caption{Galex FUV image} \label{v420fuv2}
        \end{subfigure}
        \begin{subfigure}{.49\linewidth}
                \includegraphics[width = \linewidth]{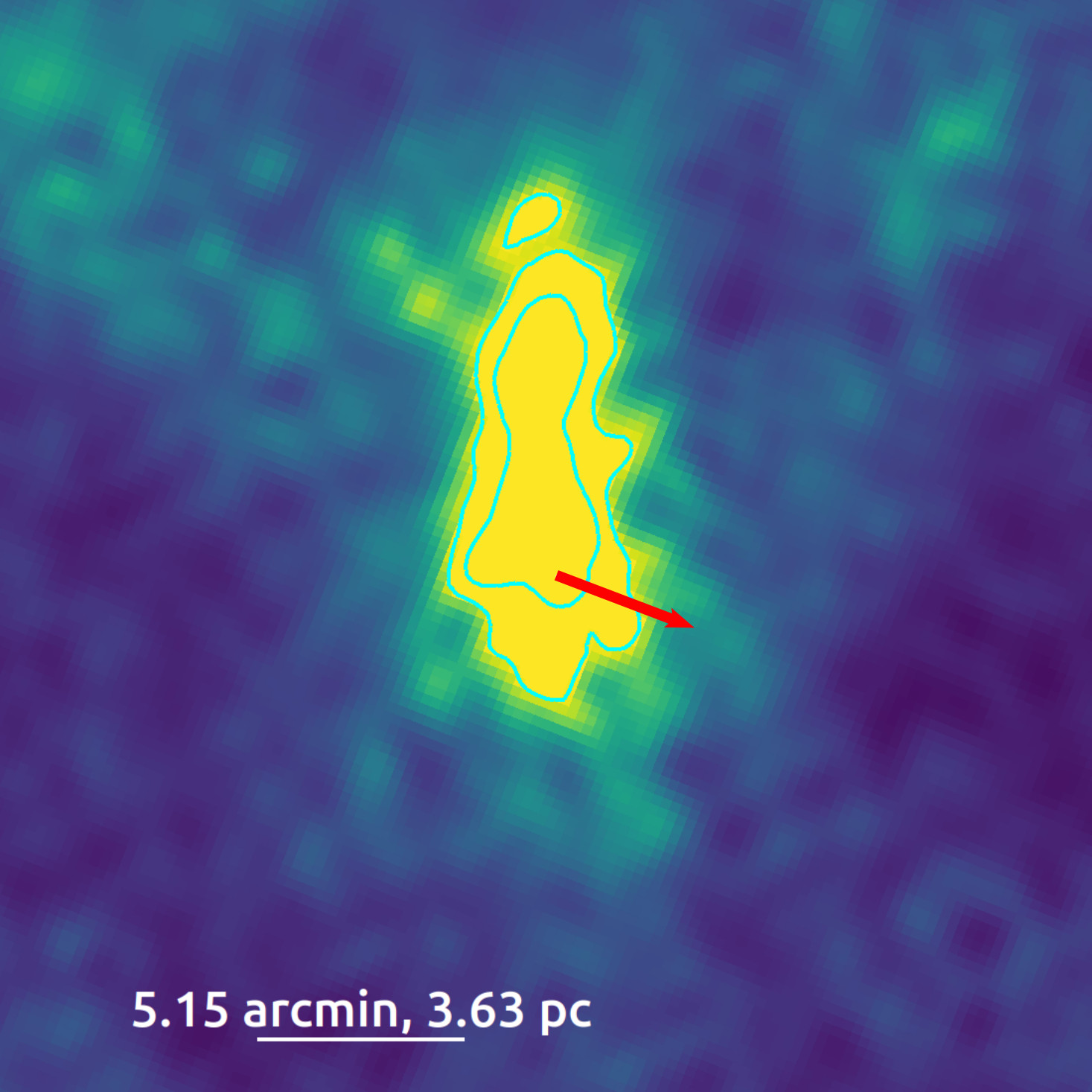}
        \caption{Akari WideS 90$\mathrm{\mu m}$ image} \label{v420fir2}
        \end{subfigure}
        \caption{Extended emission seen around \object{V420 Vul} in FUV (left) and FIR (right). The cyan contours represent the 3 and 4\,$\sigma$ levels of emission above the background in the FIR data. The arrows are the same as in Fig.~\ref{fuvall}.}\label{v420}
\end{figure}

\section{Discussion and summary} \label{s4}

\indent\indent This paper presents a short analysis of a few AGB stars that display extended emission in the FUV: five M-type, two C-type, and one S-type. The sources were identified based on the General Catalogue of Variable Stars, meaning there is a bias towards optically bright, and hence low-MLR AGB stars in this survey. In six cases the FUV emission was found to be cospatial with previous IR detections of extended emission. In three cases the FUV emission is seen in the form of circular shells centred on the stars, and in one of these cases the emission is clearly brighter in the direction of motion of the star. 

\par With the exception of two objects, no NUV emission (0.18 – 0.28 $\mu$m) is seen around the sample stars. This aspect rules out a few emission mechanisms, such as dust-scattering. One of the remaining possible processes in this case is related to shocks, most likely in the form of line emission from shock-excited $\mathrm{H_2}$, which is produced via the interaction of molecular hydrogen, originating in the cool wind, with hot electrons coming from the post-shock gas \citep[e.g.][]{martin2007, sahai2010, sahai2014}. In these instances there would be no NUV counterpart, which is observed in most of the cases presented here. With the exception of one object ($\beta$ Gru), all stars have computed space velocities of $\mathrm{40\,km\,s^{-1}}$ or more (Table~\ref{motion}), indicating that they are moving quite fast through the surrounding ISM. These high velocities are an additional argument in favour of the shock scenario, as temperatures that are high enough to shift the flux of the emission of the outflows towards the UV regime cannot otherwise be produced in the interaction and/or post-shock regions. It is also suggested that such emission would show up in clumps \citep{sanchez2015}, an aspect that is visible in some of the cases presented above. The relatively short exposure time of the UV data could be the reason for the observed clumpiness, however,  so deeper observations would be necessary to determine if this is a potential indicator of this type of FUV emission. For the cases with NUV emission or a diffuse morphology it is unclear what the origin of the UV emission is. 

\par The objects presented here increase the number of known AGB stars that display some type of UV emission. While this sample is by no means significant enough, we note that FUV emission seems to be capable of tracing large old ML structures that are otherwise too faint to be seen in the IR. This transition can already be seen in the cases of \object{R Dor} and \object{U Ant}, where the IR emission of the large shells is starting to be fairly faint, both shells having a kinematic age of more than 15\,000 years. In the case of \object{$\beta$ Gru}, whose shell has a kinematic age of 27\,500 years, there is no trace of IR emission.

\par While the UV wavelength regime has been used to observe AGB stars, extended and high-resolution observations are few in number. If the conclusions above hold, future UV studies focused on AGB stars could provide valuable information regarding stellar evolution and the enrichment of the interstellar medium.

\begin{acknowledgements}
This publication is based on data acquired with the Atacama Pathfinder Experiment (APEX) under programme ID [O-0109.F-9314A]. APEX is a collaboration between the Max-Planck-Institut fur Radioastronomie, the European Southern Observatory, and the Onsala Space Observatory. Swedish observations on APEX are supported through Swedish Research Council grant No 2017-00648.
\end{acknowledgements}

\bibliographystyle{aa_url}
\bibliography{refs}

\begin{appendix}
\section{Regions for flux density determination} \label{appA}

\FloatBarrier
\begin{figure}
    \centering
        \begin{subfigure}{.48\linewidth}
                \includegraphics[width = \linewidth]{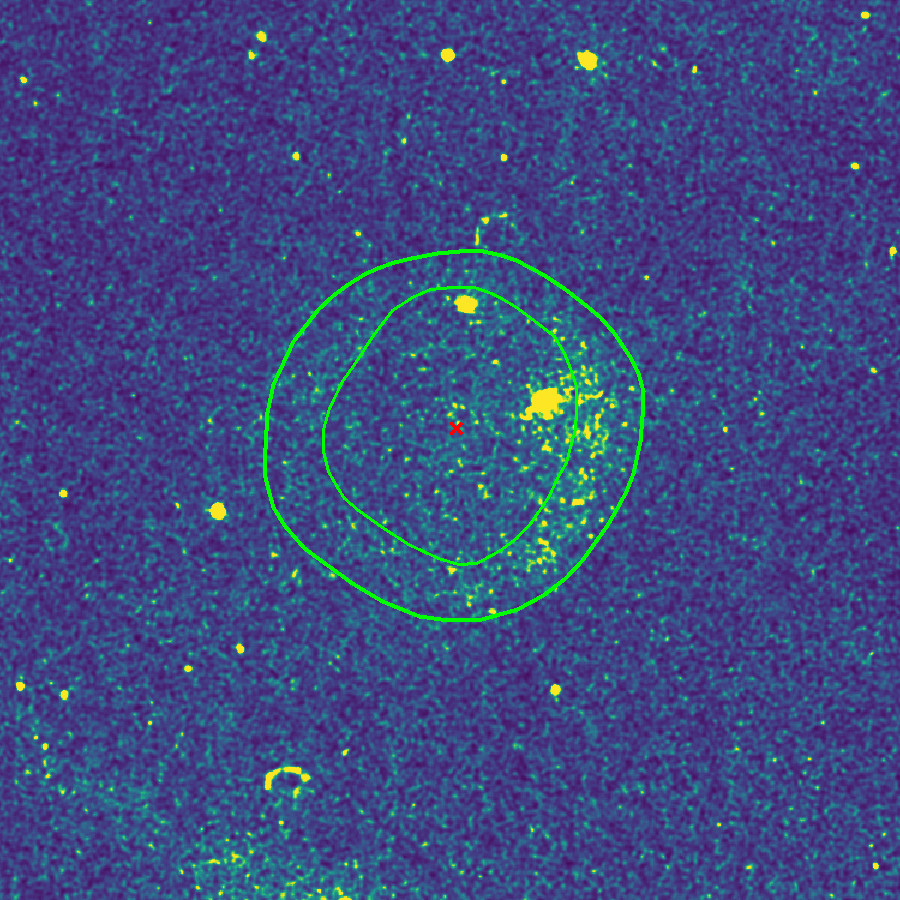}
                \caption{\object{U Ant}}
        \end{subfigure}
        \begin{subfigure}{.48\linewidth}
                \includegraphics[width = \linewidth]{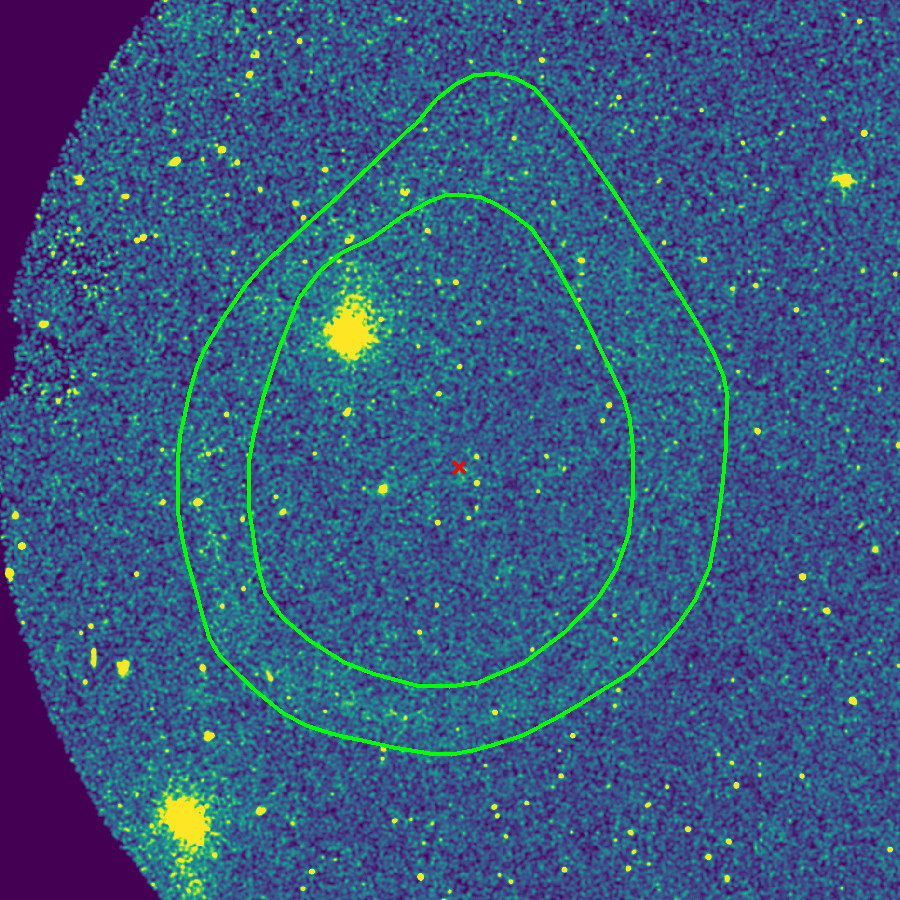}
                \caption{\object{R Dor}}
        \end{subfigure}
    \ContinuedFloat
        \begin{subfigure}{.48\linewidth}
                \includegraphics[width = \linewidth]{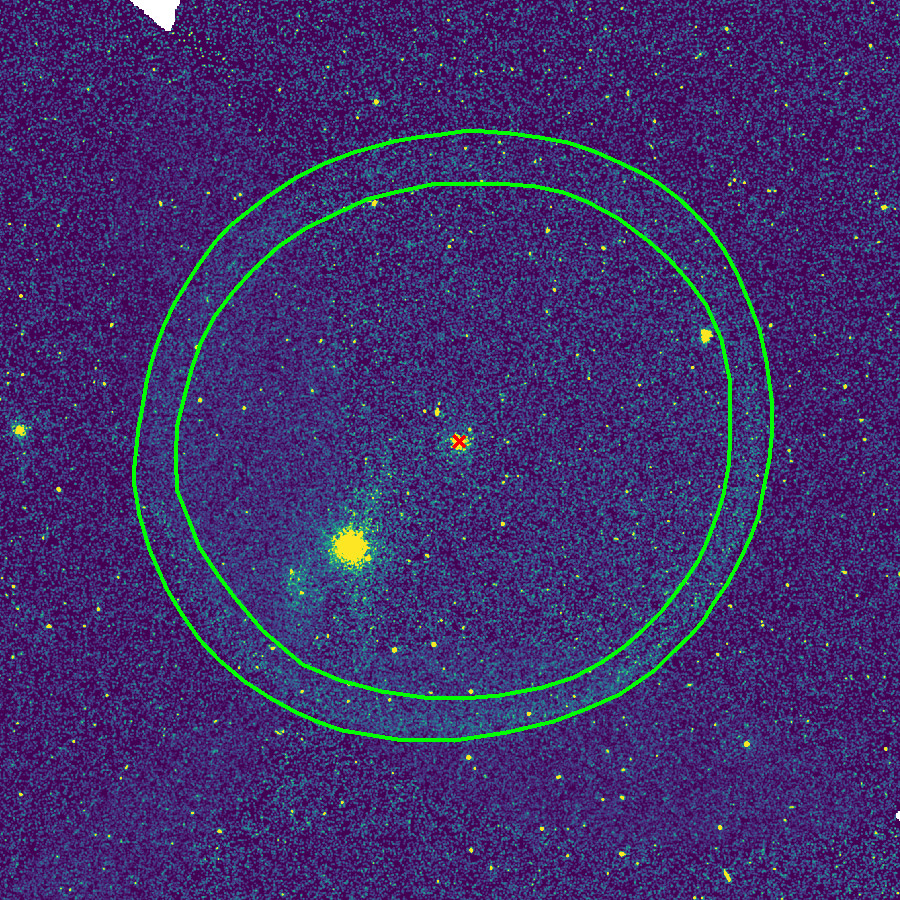}
                \caption{\object{$\beta$ Gru}}
        \end{subfigure}
        \begin{subfigure}{.48\linewidth}
                \includegraphics[width = \linewidth]{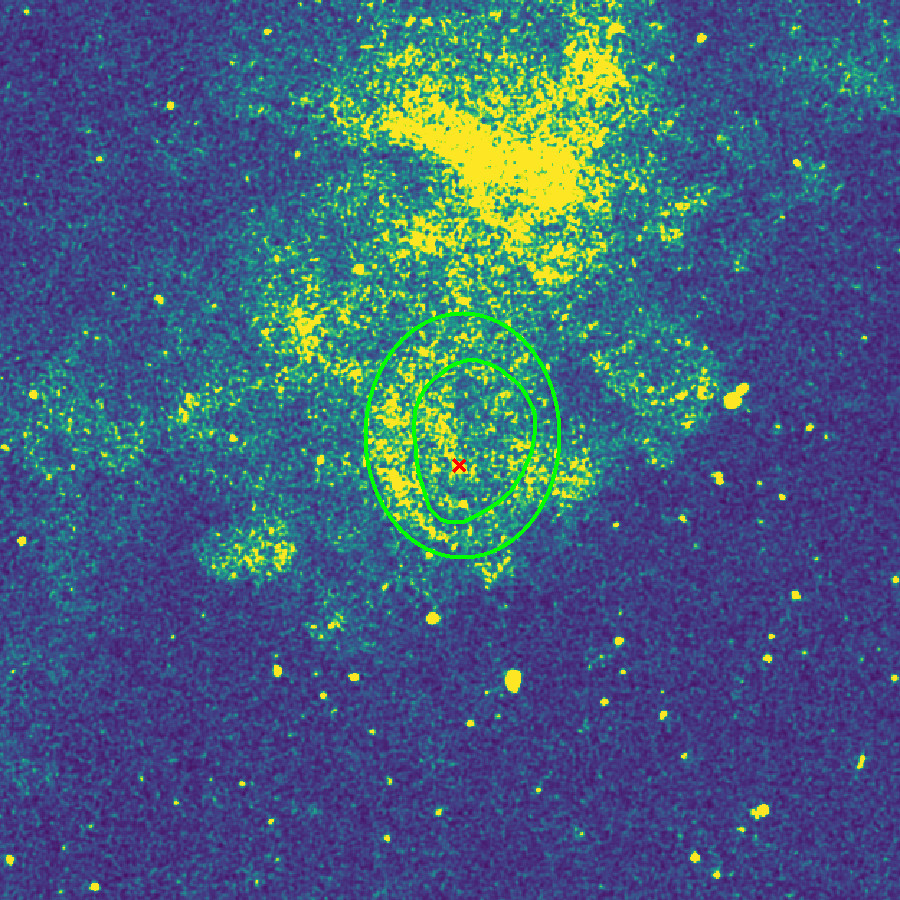}
                \caption{\object{W Hya}}
        \end{subfigure}
    \ContinuedFloat
        \begin{subfigure}{.48\linewidth}
                \includegraphics[width = \linewidth]{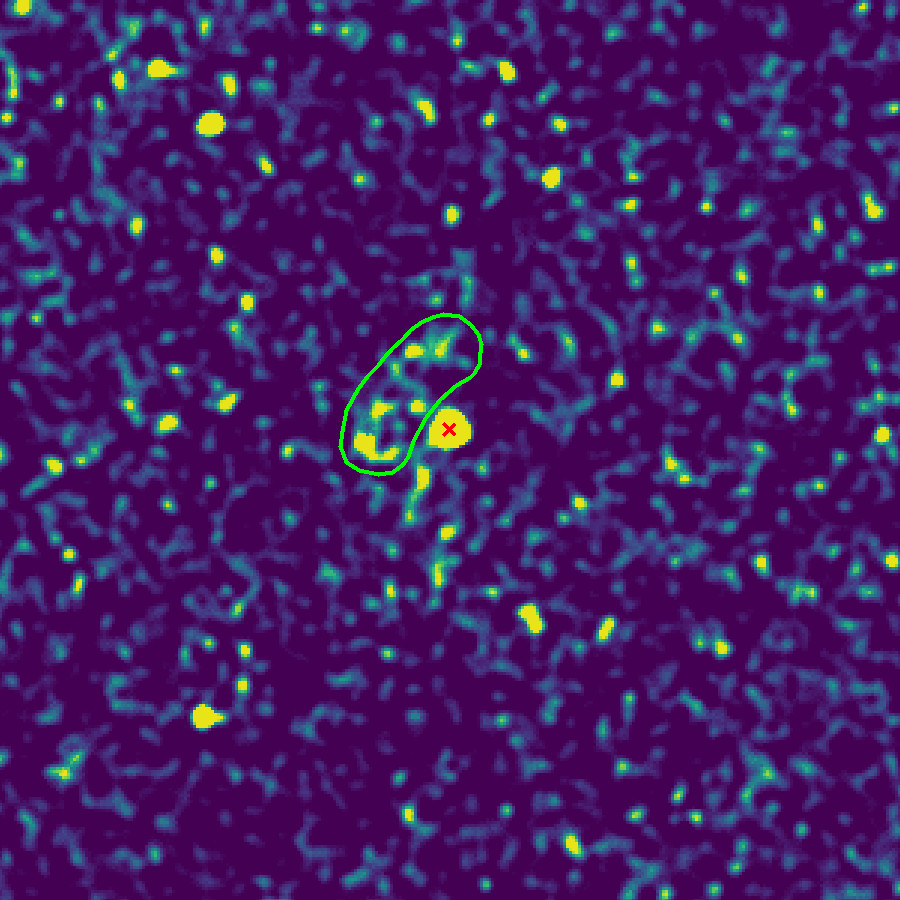}
                \caption{\object{TV Psc}}
        \end{subfigure}
        \begin{subfigure}{.48\linewidth}
                \includegraphics[width = \linewidth]{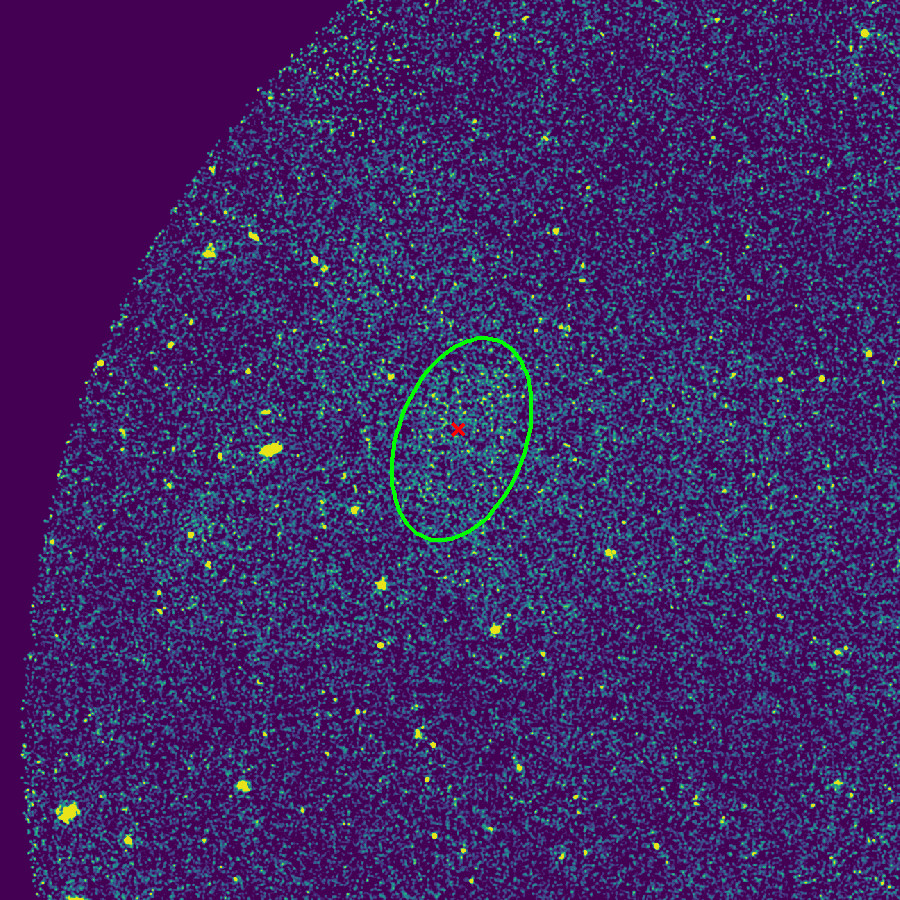}
                \caption{\object{RZ Sgr}}
        \end{subfigure}
    \ContinuedFloat
        \begin{subfigure}{.48\linewidth}
                \includegraphics[width = \linewidth]{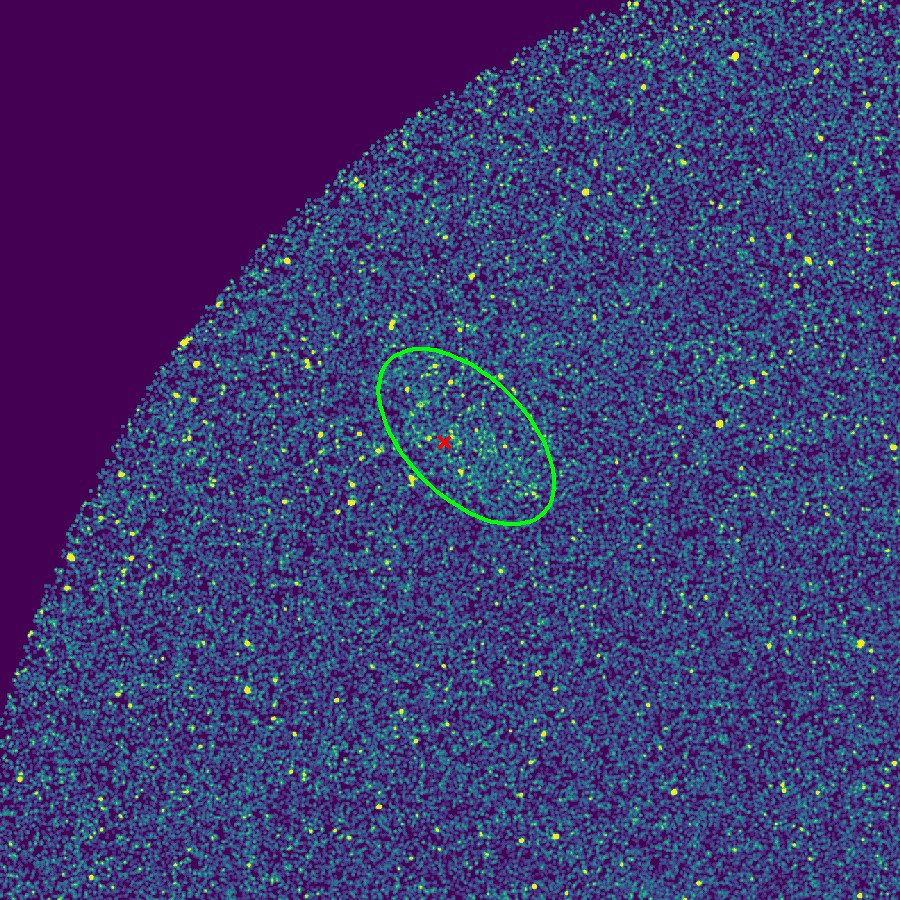}
                \caption{\object{DM Tuc}}
        \end{subfigure}
        \begin{subfigure}{.48\linewidth}
                \includegraphics[width = \linewidth]{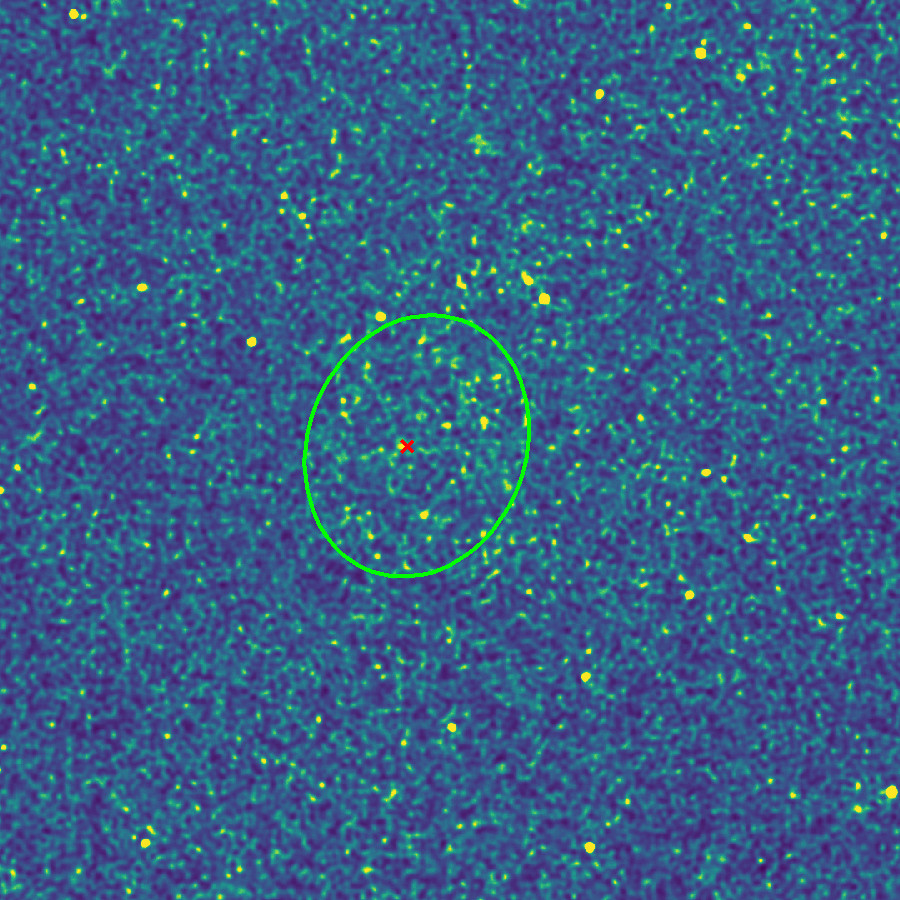}
                \caption{\object{V420 Vul}}
        \end{subfigure}
\caption{Regions used for flux density estimates. In each panel the red cross represents the location of the star.} \label{fuvregs} 
\end{figure}

\indent\indent The data used here are pipeline-calibrated data, using the \href{http://www.galex.caltech.edu/wiki/Public:Documentation/Chapter_8}{GR7 calibration process} for U Ant and the \href{https://galex.stsci.edu/Doc/GI_Doc_Ops7.pdf}{GR6 calibration process} for all other objects. Following a by-eye inspection of the FUV images, the regions were manually defined to estimate the FUV flux densities of the extended emission for each object. In the cases where the emission is diffuse, only one region was defined. For those cases where the extended structures were more pronounced, two regions were defined. The flux densities measured for the latter cases are for the area contained between the two green shapes (see Fig.~\ref{fuvregs}). A small region placed in an area of minimal emission was used in each case to determine the detector background, which was then subtracted from the obtained flux densities, as a simple means of background correction. The flux densities were obtained by converting the photon counts into  janskys via the following relation: \[\mathrm{F_{FUV}\,[Jy] = 1.115 \times 10^{-4} \times cps\,[counts/second/pixel]. }\]

\end{appendix}

\end{document}